\documentclass[a4paper,11pt]{article}
\usepackage{jinstpub} % for details on the use of the package, please see the JINST-author-manual
\usepackage{lineno}
\usepackage{lmodern}  
\usepackage{graphicx}
\usepackage{subcaption}
\usepackage{upgreek,xspace}

\usepackage{acro}

\DeclareAcronym{sd}{
	short = \textit{SD},
	long  = \textit{Structured Detection}
 }

\DeclareAcronym{si}{
	short = \textit{SI},
	long  = \textit{Structured Illumination}
}

\DeclareAcronym{dmd}{
	short = \textit{DMD},
	long  = \textit{Digital Mirror Device}
}

\DeclareAcronym{spi}{
	short = \textit{SPI},
	long  = \textit{Single-Pixel Imaging}
}

\DeclareAcronym{ccd}{
	short = \textit{CCD},
	long  = \textit{Charge-Coupled Device}
}

\DeclareAcronym{cmos}{
	short = \textit{CMOS},
	long  = \textit{Complementary Metal-Oxide-Semiconductor}
}

\DeclareAcronym{tval3}{
	short = \textit{TVAL3},
	long  = \textit{Total Variation Minimization by Augmented Lagrangian and Alternating Direction Algorithm}
}

\DeclareAcronym{nesta}{
	short = \textit{NESTA},
	long  = \textit{Nesterov's Algorithm}
}

\DeclareAcronym{pla}{
	short = \textit{PLA},
	long  = \textit{Polylactic Acid Filament}
}

\DeclareAcronym{adc}{
	short = \textit{ADC},
	long  = \textit{Analog-to-Digital Converter}
}

\DeclareAcronym{tpc}{
	short = \textit{TPC},
	long  = \textit{Time Projection Chamber}
}

\DeclareAcronym{csa}{
	short = \textit{CSA},
	long  = \textit{Charge-Sensitive Amplifier}
}

\DeclareAcronym{gem}{
	short = \textit{GEM},
	long  = \textit{Gas Electron Multiplier}
}

\DeclareAcronym{cs}{
	short = \textit{CS},
	long  = \textit{Compressive Sensing}
}

\DeclareAcronym{gpsc}{
	short = \textit{GPSC},
	long  = \textit{Gas Proportional Scintillation Counter}
}

\DeclareAcronym{pic}{
	short = \textit{PIC},
	long  = \textit{Proportional Ionization Counter}
}

\DeclareAcronym{mwpc}{
	short = \textit{MWPC},
	long  = \textit{Multiwire Proportional Chamber}
}

\DeclareAcronym{mpgd}{
	short = \textit{MPGD},
	long  = \textit{Micropattern Gaseous Detector}
}

\DeclareAcronym{tv}{
	short = \textit{TV},
	long  = \textit{Total Variation}
}

\DeclareAcronym{fwhm}{
	short = \textit{FWHM},
	long  = \textit{Full Width at Half Maximum}
}

\DeclareAcronym{cnr}{
	short = \textit{CNR},
	long  = \textit{Contrast-to-Noise Ratio}
}

\DeclareAcronym{daq}{
	short = \textit{DAQ},
	long  = \textit{Data Acquisition}
}

\title{X-ray Single-Pixel Imaging with MPGD-based detectors}

\author[a,b]{M. Sim\~oes}
\author[a]{P. Vaz}
\author[b,1,2]{and A.\,F.\,V. Cortez\note{Corresponding author.}\note{Present affiliation: AstroCeNT, Nicolaus Copernicus Astronomical Center of the Polish Academy of Sciences, ul.~Rektorska 4, 00-614 Warsaw, Poland}}
\affiliation[a]{LIBPhys-UC, Department of Physics, Rua Larga, University of Coimbra, 3004-516 Coimbra, Portugal}
\affiliation[b]{Institute of Experimental and Applied Physics, Czech Technical University in Prague, Husova 240/5, 110 00 Prague, Czech Republic}
\emailAdd{acortez@camk.edu.pl}

\abstract{X-ray imaging is an invaluable tool for noninvasive analysis in many fields ranging from basic science to medicine and security. The development of low-dose large area imaging solutions still represents an important challenge for various applications.

One solution to the imaging of large areas lies in the development of novel computational imaging systems that can overcome the limitations imposed by hardware, relying instead on numerical processing power. The single-pixel detector, depending on the application, may offer a competitive edge over conventional cameras (being a cheaper alternative to the multi-pixelated solutions). In addition, the single-pixel detector can be used to achieve improved detection efficiency, faster timing response, and good spatial resolution with low radiation dose. Moreover, this technique enables detectors to image through diffuse mediums, increasing the image quality at significant depths, solving the depth penetration issues of other imaging methods.
In this work, we explore the application of single-pixel imaging techniques to produce two-dimensional images with high temporal resolution (only if the process is reproducible and the conditions are static), using only a single detector (bucket detector). The setup, based on the application of Hadamard patterns, showed promising results, proving the ability of the system to acquire images using thin PLA based masks (up to 5 mm thickness). Both simulation, using GEANT 4, and experimental setup, based on a time projection chamber (TPC), used in this work to demonstrate this technique will be reported here along with the first results.}


\arxivnumber{}

\begin{document}
\maketitle
\flushbottom

\section{Introduction}\label{sec:intro}
Due to their nature, X-rays are an invaluable tool for noninvasive analysis in many fields ranging from basic science to medicine, engineering and security \cite{xrayimagingintro, xrayct}. 

A new X-ray imaging technique, cost-efficient, with improved detection efficiency, faster timing response, and good spatial resolution with low radiation dose is presented in this work. The techniques explored for X-ray imaging in this work is based on the so-called \ac{spi}. This technique relies on numerical processing power to achieve two dimensional images, instead of complex hardware. By using a single-pixel detector and a light/radiation modulator to project a set of patterns, \ac{spi} presents itself as a cost-effective option to multi-pixelated ones \cite{vaz2022re}. In addition, the single-pixel detector can be used to achieve improved geometry detector acceptance, given the large one pixel area of detection, and removing the non-active areas between pixels. In certain conditions, it can also obtain faster timing response \cite{santos2021compressive} when compared with frame-based readout \cite{laranja}. Moreover, this technique enables detectors to image through diffuse media, since the detector integrates all the transmitted radiation for each mask. Each mask measurement suffers an equal attenuation, resulting in the preservation of the intensity relation between the masks \cite{bguerra}.   

By combining this technique with compressing sensing \ac{cs}, it is possible to further reduce the data storage and data transfer requirements, since \ac{cs} states that an image can be reconstructed with a smaller number of measurements than the amount of pixels, under certain conditions \cite{cslaranja}. This reduction in the number of measurements can allow for a reduction in radiation dose \cite{laranja}. This can be useful for remote sensing applications and all applications of hyperspectral imaging. 

\section{Single-Pixel Imaging (SPI)}\label{sec:xxx2}

\ac{spi} has its origins in the field of \ac{cs}, which is a mathematical theory that postulates that, under certain conditions, a signal or image can be reconstructed with a small number of measurements, more exactly below the Nyquist limit \cite{Vaz2020}. 

In order to produce an image, \ac{spi} setups require: a single-pixel detector, a light source and a spatial light modulator. The spatial light modulator is employed to project different pre-selected patterns onto a scene, strategically chosen from a sensing matrix, enabling effective image reconstruction. This has proven to be revolutionary, as it allows obtaining similar or higher resolution images, without the need to further increase the cost of the hardware, while making use of a sub-set of samples~\cite{candes2008introduction}. The mathematical formulation of \ac{spi} to obtain the sensing matrix will be address next. Even though \ac{cs} is more effective and important only when imaging with high resolutions \cite{vaz2022re}, in this work we opted for a \ac{spi} approach due to the limited resolution allowed by the setup. 
% Still it is important to describe mathematically both \ac{cs} and \ac{spi}. Mathematically, \ac{cs} can initially be presented as follows:
% \begin{equation} \label{eqcs}
%     x = \Psi \theta
% \end{equation}
% \noindent where $x \in \mathbb{R}^{N \times 1}$ is the discrete signal and $\theta \in \mathbb{R}^{N \times 1}$ is the sparse vector that contains the $x$ projections in the orthogonal basis, $\Psi \in \mathbb{R}^{N \times N}$ \cite{cslaranja, santos2021compressive}.

% Two important requirements in \ac{cs} are sparsity and incoherence. A signal must be sparse, \textit{i.e.}, few coefficients are able to represent the signal, and the sensing matrix, ($\Phi$), must have low coherence with $\Psi$ in order to form an $N$ resolution image with $M$ measurements, where $M < N$. In most applications, these are independent of the signal to be measured, resulting in a sampling process:
% \begin{equation} \label{cs2}
%     y = \Phi x = \Phi \Psi \theta = A \theta  
% \end{equation}

% \noindent where $y \in \mathbb{R}^{M \times 1}$ contains the sampled data projections using the sensing matrix ($\Phi$) and $A$ is the reconstruction matrix. In practical terms, $\Phi$ is projected onto the scene and $A$ is used to reconstruct the final image \cite{Vaz2020}.

\subsection{Sensing Matrix}

When designing a sensing matrix the level of incoherence with the orthonormal reconstruction basis must be taken into account. It has been proven that random patterns show the low incoherence required, but these introduce an additional degree of freedom, as they are not pre-designed matrices. 
% Other options are Fourier, wavelet and Hadamard \cite{hadavsfour}.

This work will focus on the implementation of Hadamard transform matrix, which has already been demonstrated to be suitable for \ac{spi}. The Hadamard matrix consists of a square matrix of only -1/1 values, obtained from the Walsh functions, orthogonal binary functions \cite{lopez2022efficient, Vaz2020}. This option stands out due to its simple implementation both in hardware and computationally.

The Hadamard matrices of first and second order are defined as \cite{lopez2022efficient}:

\begin{equation}
    H_1 = \mathbf{\begin{bmatrix}
    1 \end{bmatrix}}~ \text{and}~ H_2 = \mathbf{\begin{bmatrix}
    1 & 1 \\
    1 & -1
\end{bmatrix}} 
\end{equation}
From $H_2$, Natural ordering Hadamard transform matrix of order $2^k$ can be derived recursively with the following equation:

\begin{equation}
    H_{2^k} = \mathbf{\begin{bmatrix}
    H_{2^{k-1}} & H_{2^{k-1}} \\
    H_{2^{k-1}} & -H_{2^{k-1}}
\end{bmatrix}} = H_2 \otimes H_{2^{k-1}} 
\end{equation}

\noindent where k is an integer larger than 1 and $\otimes$ is the Kronecker product. The Kronecker product is an operation that combines two matrices by multiplying each element of one matrix with the entire other matrix, resulting in a larger matrix, more specifically with a size equal to product of the sizes of the two multiplied matrices \cite{kronecker}. Considering the Natural ordering Hadamard transform matrix of order 4, the process, step by step, would be the following:

\begin{equation}
    H_4 = \textcolor{red}{H_2} \otimes \textcolor{blue}{H_2}
\end{equation}
%\newpage
Then,

\begin{equation} \label{eq:h4}
    H_4 = \mathbf{\begin{bmatrix}
    1 & 1 \\
    1 & -1
\end{bmatrix}} \otimes \mathbf{\begin{bmatrix}
    1 & 1 \\
    1 & -1
\end{bmatrix}} = \mathbf{\begin{bmatrix}
    \textcolor{red}1 \times \textcolor{blue}{\mathbf{\begin{bmatrix}
    1 & 1 \\
    1 & -1
\end{bmatrix}}} & \textcolor{red}1 \times \textcolor{blue}{\mathbf{\begin{bmatrix}
    1 & 1 \\
    1 & -1
\end{bmatrix}}} \\
    \textcolor{red}1 \times \textcolor{blue}{\mathbf{\begin{bmatrix}
    1 & 1 \\
    1 & -1
\end{bmatrix}}} & \textcolor{red}{-1} \times \textcolor{blue}{\mathbf{\begin{bmatrix}
    1 & 1 \\
    1 & -1
\end{bmatrix}}}
\end{bmatrix}}
    = \mathbf{\begin{bmatrix}
    1 & 1 & 1 & 1 \\
    1 & -1 & 1 & -1 \\
    1 & 1 & -1 & -1 \\
    1 & -1 & -1 & 1 \\
\end{bmatrix}}
\end{equation}

\begin{figure}[ht]
    \centering
    \includegraphics[scale=0.2]{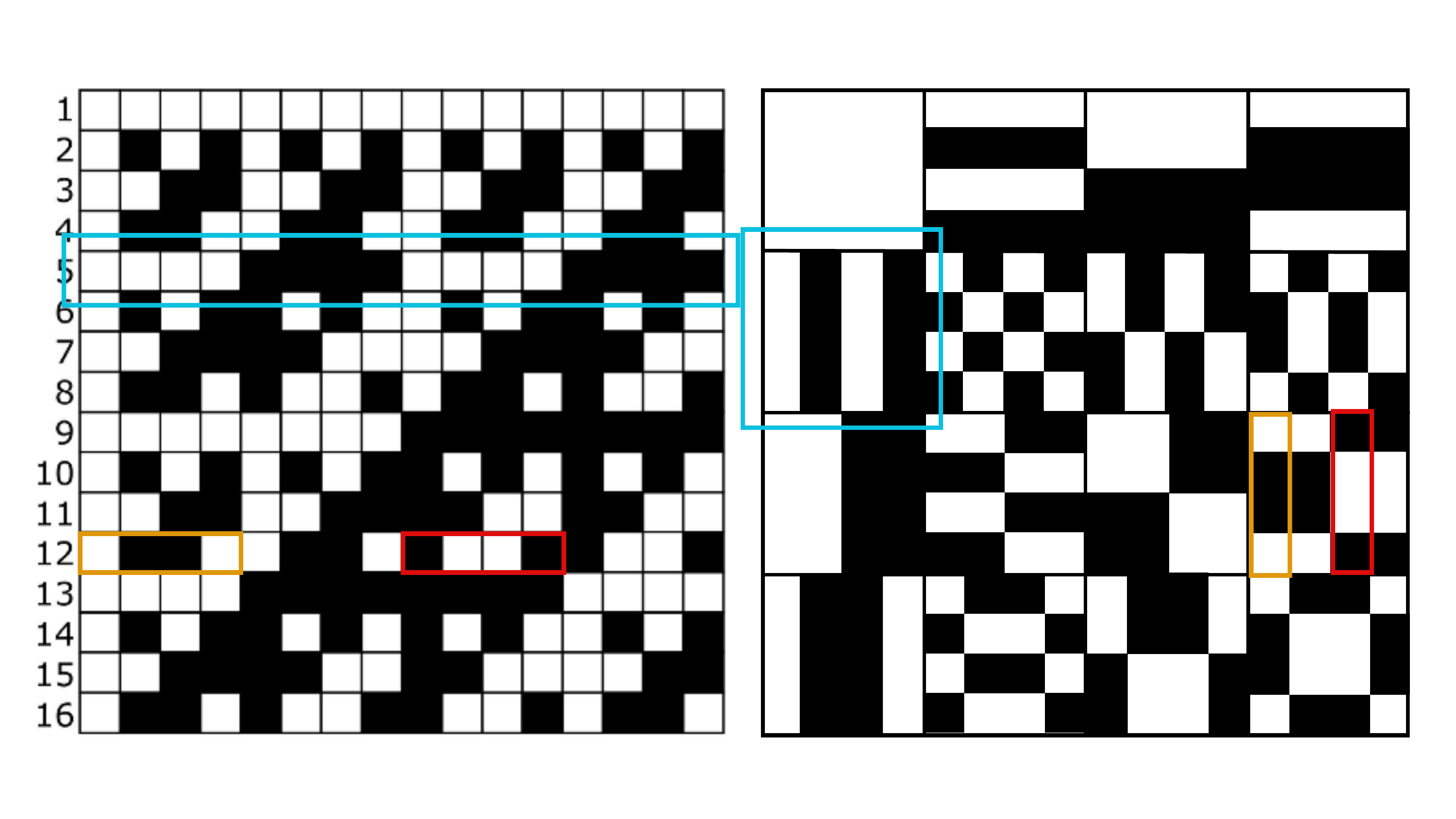}
    \caption{Hadamard matrix of order 16 in natural ordering with the corresponding reshaped patterns. Adapted from \cite{vaz2022re}.}
    \label{fig:hada}
\end{figure}

This results in a relatively simple operation to perform. Each row of the Hadamard matrix can serve as a measurement mask, by reshaping its values into a 2D array. We start by dividing each row in 4 groups of 4 values, each will correspond to a column of a measuring mask \cite{lopez2022efficient}. In Figure \ref{fig:hada}, in blue, it is observable that row 5 corresponds to mask 5. In orange and red, it is shown how a group of 4 values is transformed into a column. By performing these steps on every row, the necessary set of maks is obtained. 

So, the Hadamard matrix is used as a set of orthogonal basis patterns. By sequentially applying these measurement masks and performing the corresponding measurements, it is possible to reconstruct an image using computational algorithms.

As a way of reducing the background noise, differential Hadamard \ac{spi} has emerged. In this technique, in addition to using the set of masks mentioned, a measurement, $m_{H^{-}}$, is acquired with their inverse masks too. The value $m_{H^{-}}$, registered for each case is respectively subtracted from the value obtained with ``original'' masks, $m_{H^{+}}$, obtaining $x = m_{H^{+}} - m_{H^{-}}$.  From this point, the reconstruction is carried out \cite{lopez2022efficient}.

\subsection{Computational Algorithm} \label{chap2compal}

In cases where transformation matrices have been applied, such as the Hadamard, the simplest way to reconstruct is to use their inverse transform. 

% Considering, a column matrix $y$ of measured data,  a column matrix $x$ of target data and $H$, the Hadamard matrix, can be written as follows:

 \begin{equation} \label{eq:resHad}
     \mathbf{\begin{bmatrix}
     x_1 \\
     \vdots \\
     x_n 
     \end{bmatrix}} = H^{-1}
     \mathbf{\begin{bmatrix}
     y_1 \\
     \vdots \\
     y_n 
     \end{bmatrix}}
     =
     \frac{1}{n} 
     H
     \mathbf{\begin{bmatrix}
     y_1 \\
     \vdots \\
     y_n 
     \end{bmatrix}}
 \end{equation}

So, by transposing the Hadamard matrix, then multiplying it with the final measured values, $y$, we obtain $x$ with the $n$ final coefficients. One detail to note is that the inverse Hadamard matrix can be obtained by dividing the original Hadamard matrix by its dimension $n$, which further simplifies the computational work. By reshaping $x$ into a $\sqrt{n}\times\sqrt{n}$ matrix and displaying it as 2D array, we will form the reconstructed image \cite{hadavsfour}.

Using the Hadamard matrix obtained previously (equation \ref{eq:h4}), the respective masks and the formula \ref{eq:resHad}, a simple example for 4 pixels is given next. Considering the following:

\begin{figure}[h!]
    \centering
    \includegraphics[scale=0.3]{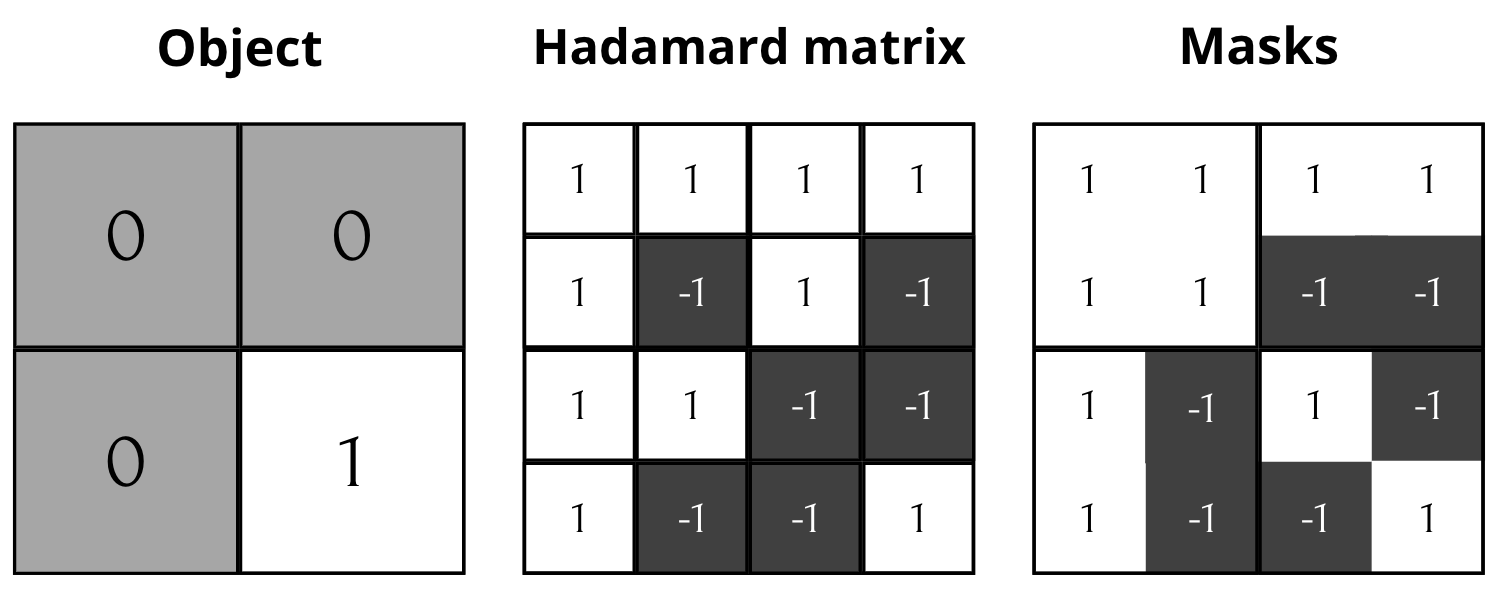}
    \caption{Elements used for the example.}
    \label{fig:example}
\end{figure}

Each mask, presented on the right in Figure \ref{fig:example}, is projected onto the object, presented on the left. Analytically, this procedure can be expressed using equation \ref{eq:cs2}:
\begin{equation} \label{eq:cs2}
    y = \Phi x  
\end{equation}
that results in the following coefficients:
\begin{equation}
 y = \mathbf{\begin{bmatrix}
    1 \\
    -1 \\
    -1 \\
    1
\end{bmatrix}}
\text{or, when reshaped,}~y = \mathbf{\begin{bmatrix}
    1 & -1 \\
    -1 & 1
\end{bmatrix}}
\end{equation}

Finally, by applying the reconstruction formula (equation \ref{eq:resHad}) it is possible to obtain the object image as follows.

\begin{equation}
    x = H^{-1} \times y = 
    \frac{1}{4} \mathbf{\begin{bmatrix}
    1 & 1 & 1 & 1 \\
    1 & -1 & 1 & -1 \\
    1 & 1 & -1 & -1 \\
    1 & -1 & -1 & 1 \\
\end{bmatrix}} \times \mathbf{\begin{bmatrix}
    1 \\
    -1 \\
    -1 \\
    1
\end{bmatrix}} = \mathbf{\begin{bmatrix}
    0 \\
    0 \\
    0 \\
    1
\end{bmatrix}} \text{or}~x = \mathbf{\begin{bmatrix}
    0 & 0 \\
    0 & 1
\end{bmatrix}}
\end{equation}

Other more complex and more effective algorithms can be used, both as a post-processing step after the reconstruction or as independent image reconstruction algorithms. Some of these include the use of optimization algorithms to reconstruct images with the minimum number of measurements, such as the \ac{tval3} and the \ac{nesta} \cite{vaz2022re}.

\section{MPGD-based Single-Pixel Imaging setup}
%In the present section we will address the main considerations when developing a \ac{spi} system, from the detector setup to definition/development of the 3D-printed set of masks, taking into account both simulation and experimental results, including a full characterization of the detector.

%For the definition of the experimental setup, we considered both the requirements from the point of view of the detector, in our case, a triple \ac{gem} detector and the standard \ac{spi} setup, bearing in mind the fact that we used X-ray radiation.  

A single pixel camera based on \ac{sd} was developed to test the concept of X-ray \ac{spi}. The designed camera was composed by three main components/parts: the set of masks that contain the different patterns that will allow reconstruct 2D images, a triple \ac{gem} detector that will enable the detection of the X-rays, with the electronics associated, and an X-Ray tube that will be used as an X-ray source in this work, as shown in figure~\ref{fig:setupfim}. 
These triple-\ac{gem} detectors are typically formed by a cathode, a stack of three \ac{gem} structures and an anode placed inside a gas vessel, with the \ac{gem} structures consisting of a thin, metal-clad polymer foil, chemically etched with a high density of holes (typically 10 to 100 per mm$^2$). 

\subsection{Working principle and Experimental setup}

Upon interacting in the absorption region (between the cathode and triple-\ac{gem} stack), low-energy X-ray photons will originate a primary charge cloud (proportional to its energy). Due to the action of an external uniform electric field, the primary electrons will be drifted towards the triple-\ac{gem}, and then focused into the holes, where a sufficiently strong electric field will produce a charge avalanche at each amplification stage, each corresponding to an individual \ac{gem}, with the resulting charge being proportional to the primary charge and consequently to the energy of the incident ionizing radiation. 

\begin{figure}[h!]
    \centering
    \includegraphics[scale=0.5]{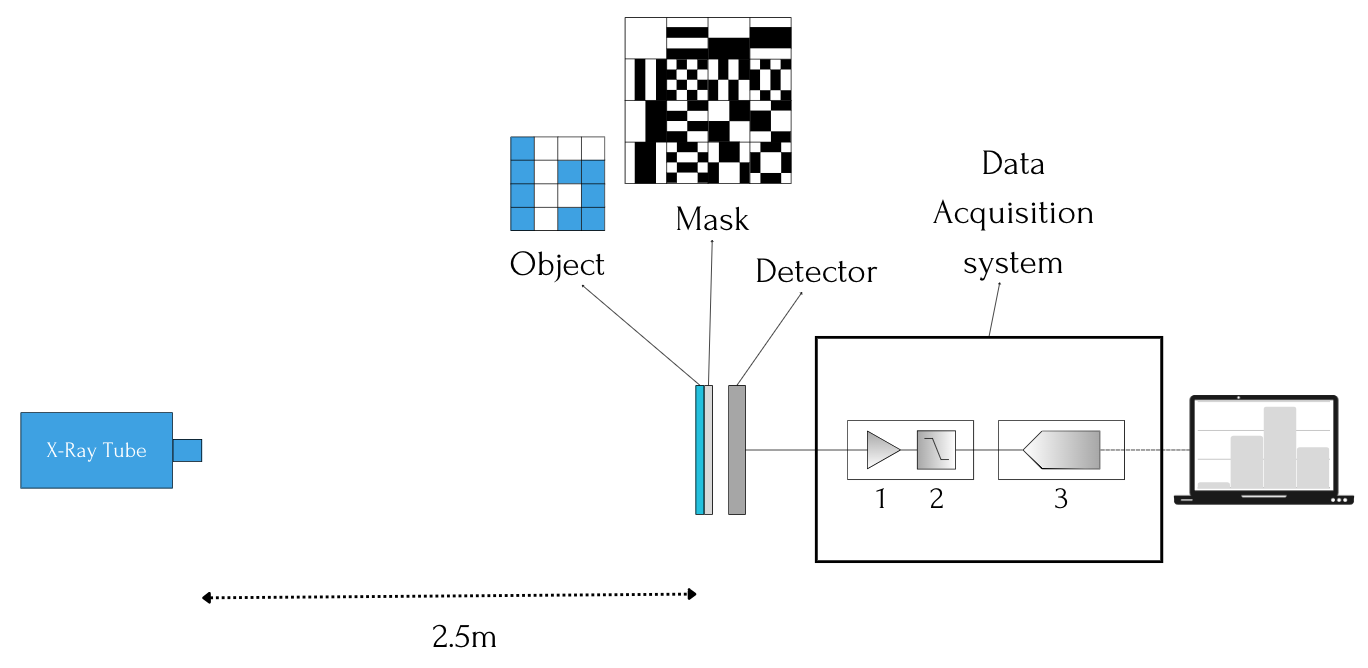}
    \caption{Schematics of the experimental setup formed by an X-ray tube, an object (letter F), the Hadammard mask and the X-ray detector (MPGD-based) followed by the \ac{daq} system (\ac{csa} (1), a shaping amplifier (2) and an \ac{adc} (3)) connected to a computer.}
    \label{fig:setupfim}
\end{figure}

By counting the number of events (X-ray photons) and determining their energy it is possible to obtain an optimized intensity profile of the irradiated structures. This intensity profile, related to the attenuation of X-ray beam caused by the different masks and object combinations, can then be used to reconstruct the original image of the object as described earlier.
Figure~\ref{fig:setupfim} shows the MPGD-based SPI setup used.

\newpage

\begin{figure}[h!]
    \centering    
    \includegraphics[scale=0.15]{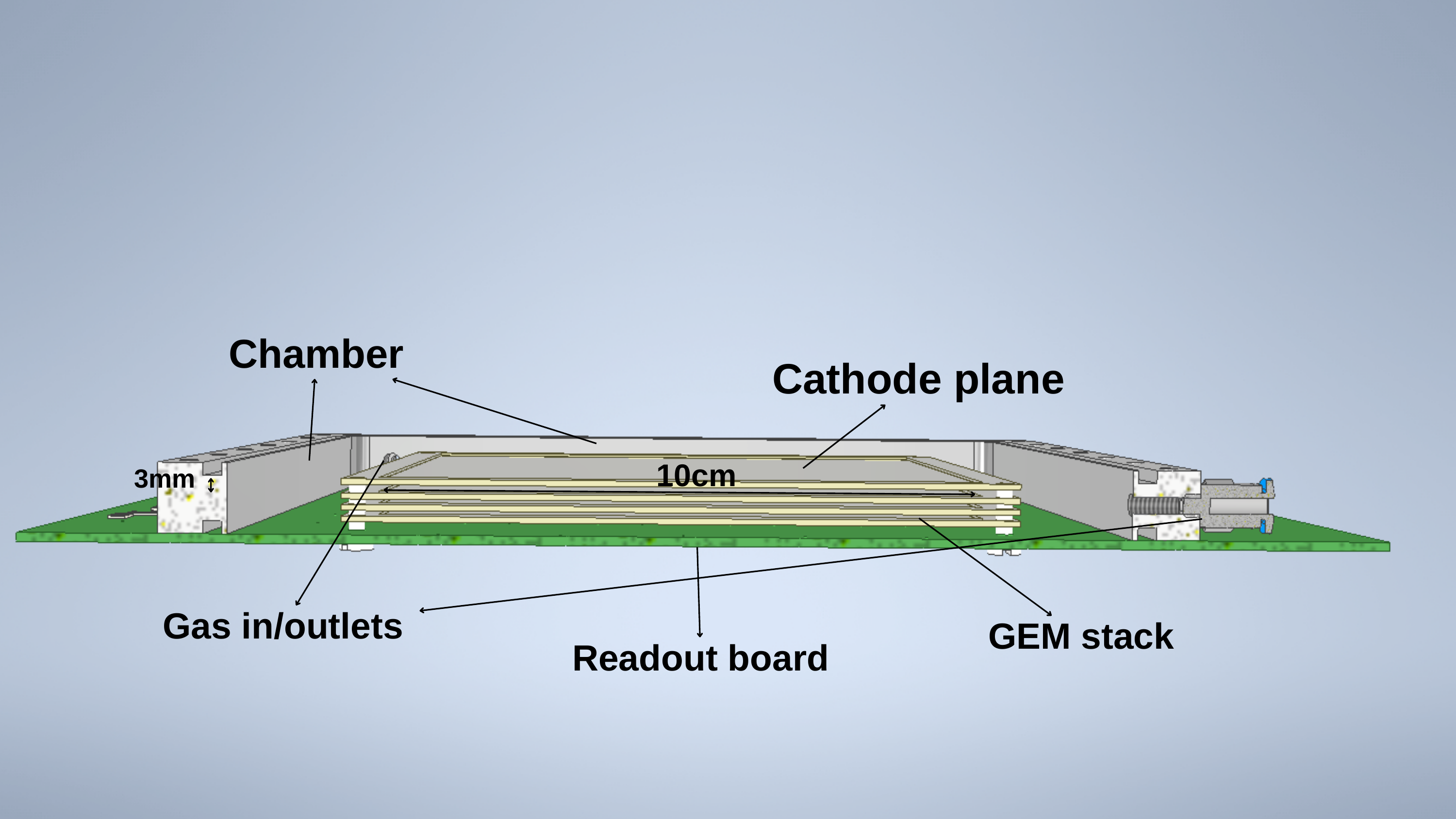}
    \hspace{1cm}
    \includegraphics[scale=0.11]{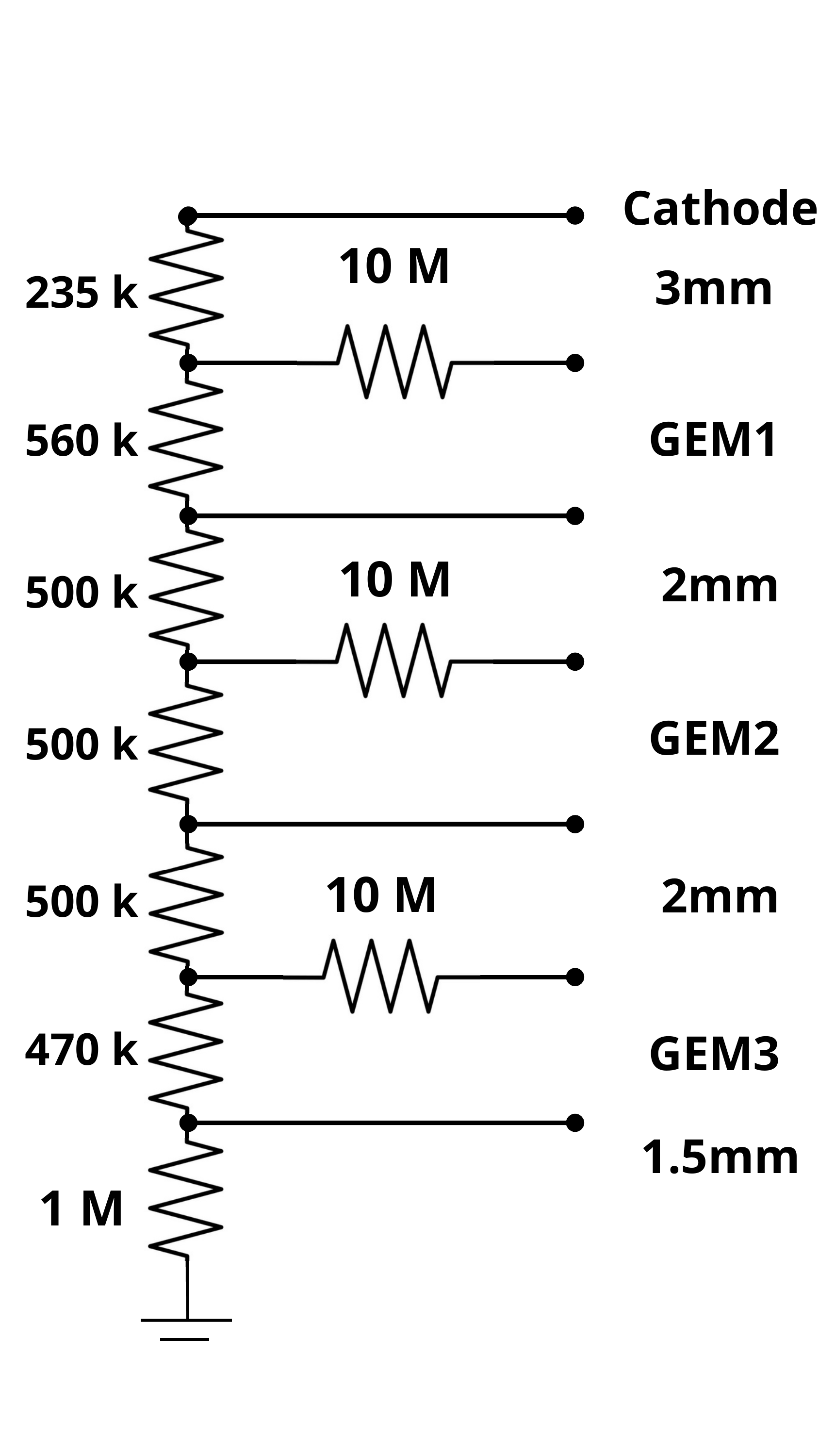}
    \caption{Detector schematics (left) and correspondent voltage divider used in this work (right).}
    \label{fig:tpcscheme}   
\end{figure}

In this work a X-ray tube (Mini-X X-Ray Tube from AMPTEK) was used as X-ray source. The X-ray source was placed at a significant distance from the remaining setup elements to ensure that the photons have parallel trajectories and impinge the different surfaces (object/mask and detector window) perpendicularly, ensuring an uniform coverage. In this case, both the number and energy of photons reaching the active volume of detection will depend on the distance and materials/structures (namely atomic number, density, thickness, etc.) encountered on their trajectory. 
The detector used was a 10 $\times$ 10 cm$^2$ triple-\ac{gem} detector as presented in Figure \ref{fig:tpcscheme}, with a 50 $\mu$m thick window made of kapton. Each \ac{gem} foil has standard dimensions, with 70 $\mu$m diameter holes and 140 $\mu$m hole-pitch, arranged in an hexagonal pattern. The detector has a 3 mm absorption/drift gap, 2 mm transfer gaps and 1.5 mm induction gap. The chamber is operated with Ar-CO$_2$ (70/30) at a rate of 6 L/h through the gas in/outlets. To bias the entire detector a fixed resistance approach (using a voltage divider) controlled by means of current was used (see Figure \ref{fig:tpcscheme} - right).

In this work, a typical effective gain of approximately $10^4$ was used, corresponding to voltages across the GEM foils of around 400 V, a typical voltage used in such detectors \cite{gemsauli}. By maintaining a slightly higher voltage in the first \ac{gem}, allows to obtain a higher initial charge gain larger, that is reflected in a better energy resolution. As the consequences of the statistical fluctuations introduced by the other \ac{gem} foils are reduced. On the other hand, the last \ac{gem} foil, which is more susceptible to electrical discharges because of the much larger number of electrons has a slightly lower voltage, that can improve the operational stability of the detector \cite{altunbas2002construction}. Considering the voltage divider presented on Figure \ref{fig:tpcscheme} (right), 2830 V corresponds to approximately 760$\mu$A through the voltage divider. That results in 600 V/cm drift field, 2 kV/cm field in the transfer gaps and  5 kV/cm field in the induction gap. 

To acquire the signal a ORTEC 142B (charge pre-amplifier) was used along with a Canberra Model 2111 (shaping amplifier), following which the signal is digitized using a Canberra ADC 8715 module which is connected to a computer for further data processing and analysis. For the biasing of the different internal structures of the detector a CAEN Module N8034 supply was selected.

\subsection{Energy Resolution}

\begin{figure}[h!]
    \centering    
    \includegraphics[width=.7\textwidth]{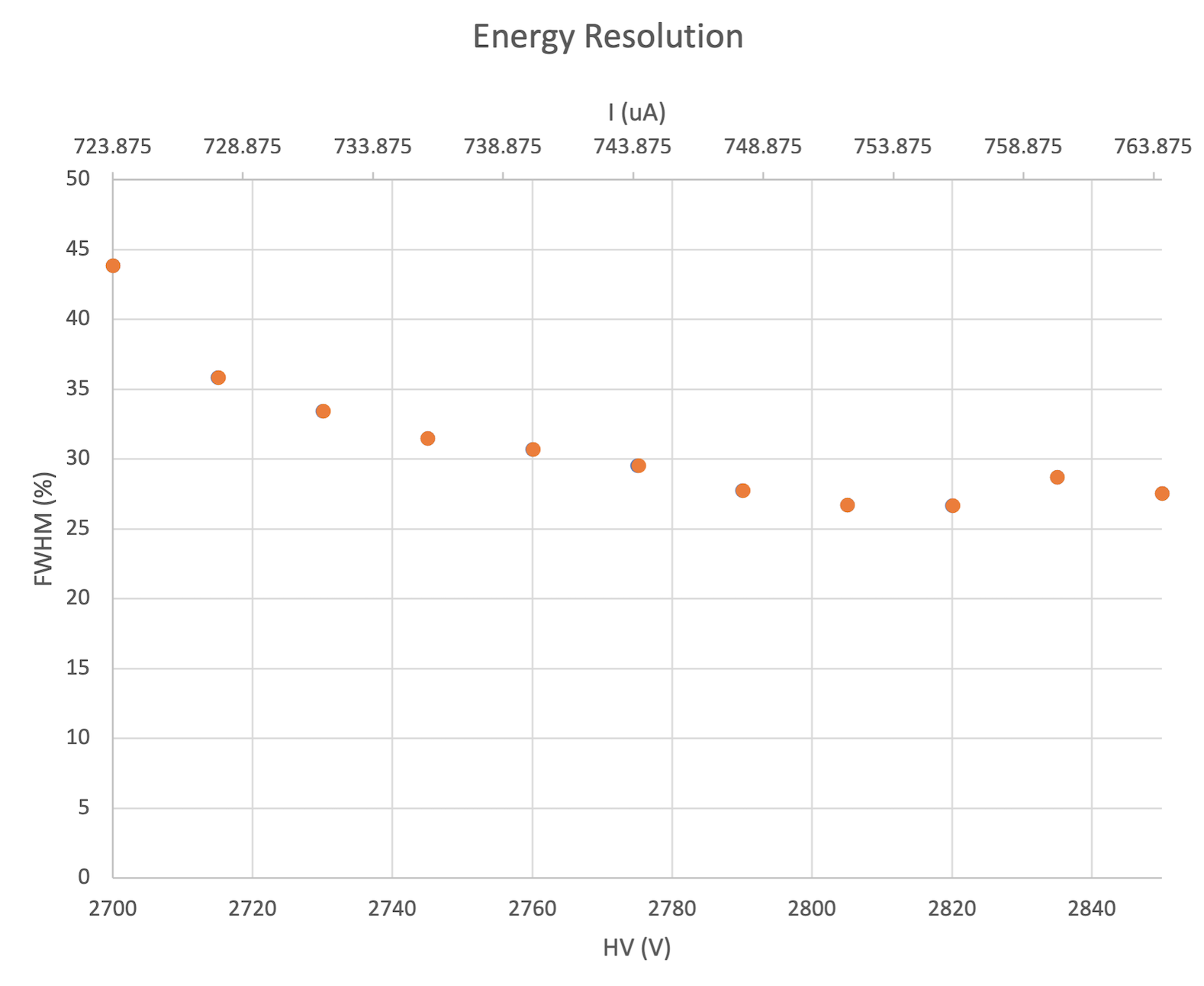}
    \caption{Energy resolution of the detector (FWHM) as a function of the total biasing voltage (V) and respective current ($\mu$A).}
    \label{fig:energyresolution}
\end{figure}

The energy resolution was determined for different \ac{gem} biasing voltages, with a $^{55}\text{Fe}$ source (5.9 keV), by calculating the \ac{fwhm}, where $\sigma$ is obtained from the Gaussian fit parameters. The values obtained are presented in Figure \ref{fig:energyresolution}. 

% Is important to note that this study was done mainly in order to characterize the detector, because for this work only the number of events detected, and not their energy, was necessary. We can see that, initially the energy resolution improves, since the gain of the detector increases and, consequently there is an improvement in signal-to-noise ratio. At the end, since the drift field increases, primary electrons start to get lost in the top of the first \ac{gem} surface.
%\newpage
\subsection{Detector Gain} \label{secdetgain}

\begin{figure}[h!]
    \centering
    \includegraphics[width=.7\textwidth]{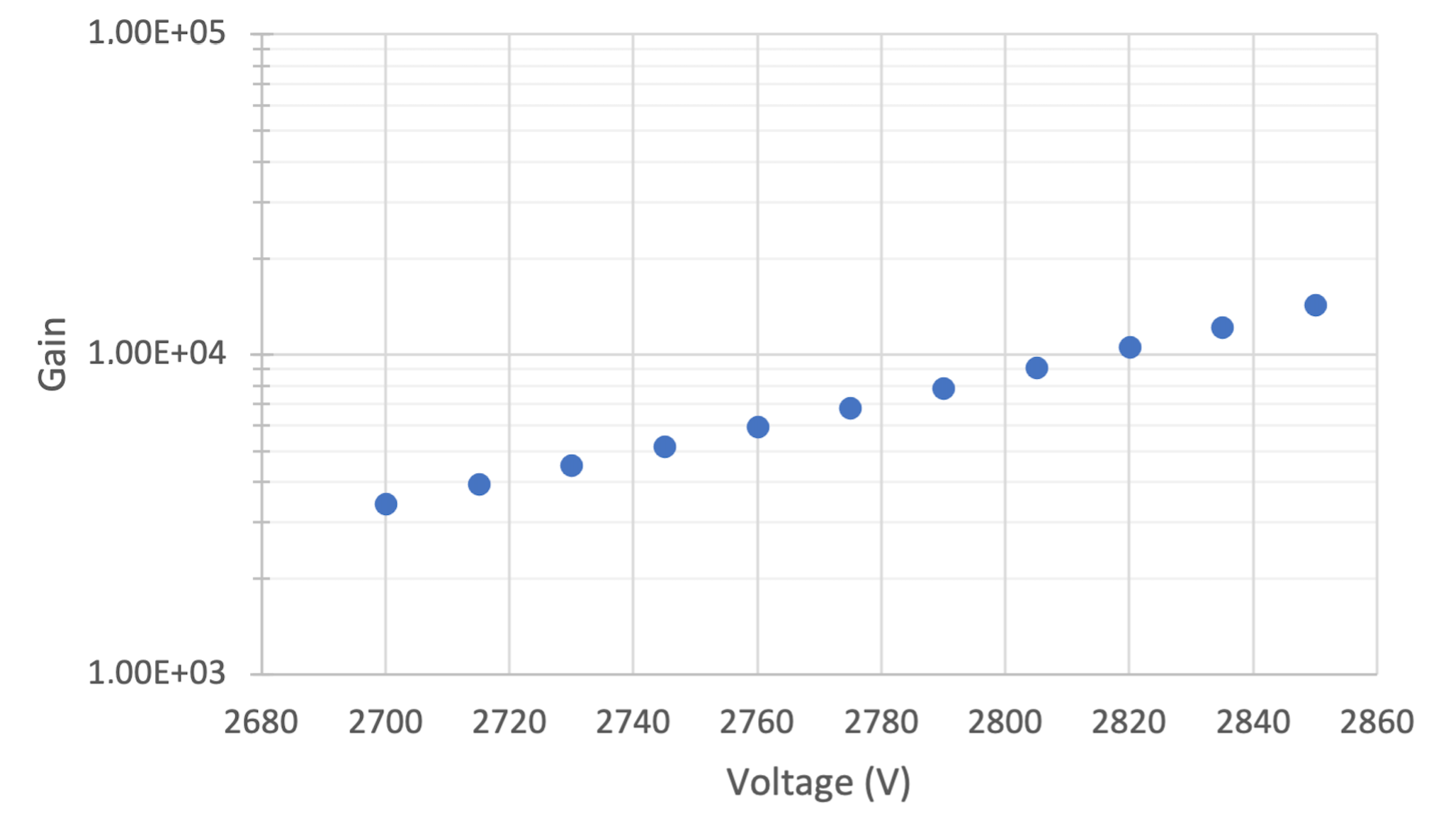}
    \caption{Logarithmic plot of the gain as a function of the applied voltage for the $^{55}\text{Fe}$ source.}
    \label{fig:gainvoltage}
\end{figure}

In addition to the energy resolution, the detector gain was determined experimentally using with a $^{55}\text{Fe}$ source, and calculated as follows:

\begin{equation} \label{eq:gain}
    G = \frac{Q_f}{Q_p} = \frac{f_{cal}(M)}{q_e \times E_{photon}/\text{W-value}}
\end{equation}

\noindent where $Q_f$ is the charge detected, $Q_p$ is the primary charge, $M$ is the mean \ac{adc} value of the $^{55}\text{Fe}$ peak, $f_{cal}$ is the electronics calibration function, $q_e$ is the electron charge, $E_{photon}$ is the $^{55}\text{Fe}$ photon energy and W-value is the mean energy necessary to create an ion pair in Ar-CO$_2$ ($70/30$). The electronics calibration function was obtained by using a waveform generator connected to inject a pulse signal and determining the signal voltage amplitude before and after amplification. The charge was determined for each input voltage, using the preamplifier's capacitance, which enabled the determination of the peak centroid in the \ac{adc}. With this information, it is possible to plot the input charge as function of the mean \ac{adc} value. The results obtained are shown in Figure \ref{fig:gainvoltage}, where an exponential trend is observed, as expected.

\newpage
\subsection{Detector Uniformity} \label{secdetuni}

 %\noindent \textbf{Uniformity} - 
Several reasons can lead to a non-uniform response from the detectors. The reasons can arise from localized defects to non-conformity in the holes and \ac{gem} foils \cite{PATRA201725}. In order to compensate for non-uniformity response of the detector, a map of the detector uniformity was obtained, using as reference the peak position. Using \ac{pla} 5 mm-thick masks with a single opening (25 $\times$ 25 mm$^2$ open area corresponding the area of the matrix pixel area) and the same total area of the masks used it is possible to evaluate individually each pixel.

% \begin{figure}[ht]
%     \centering
%     \includegraphics[scale=0.4]{figures/onepixel.jpeg}
%     \caption{Masks design for the uniformity test.}
%     \label{fig:1pixel}
% \end{figure}

\begin{figure}[h!]
    
         \centering
         \includegraphics[scale=0.38]{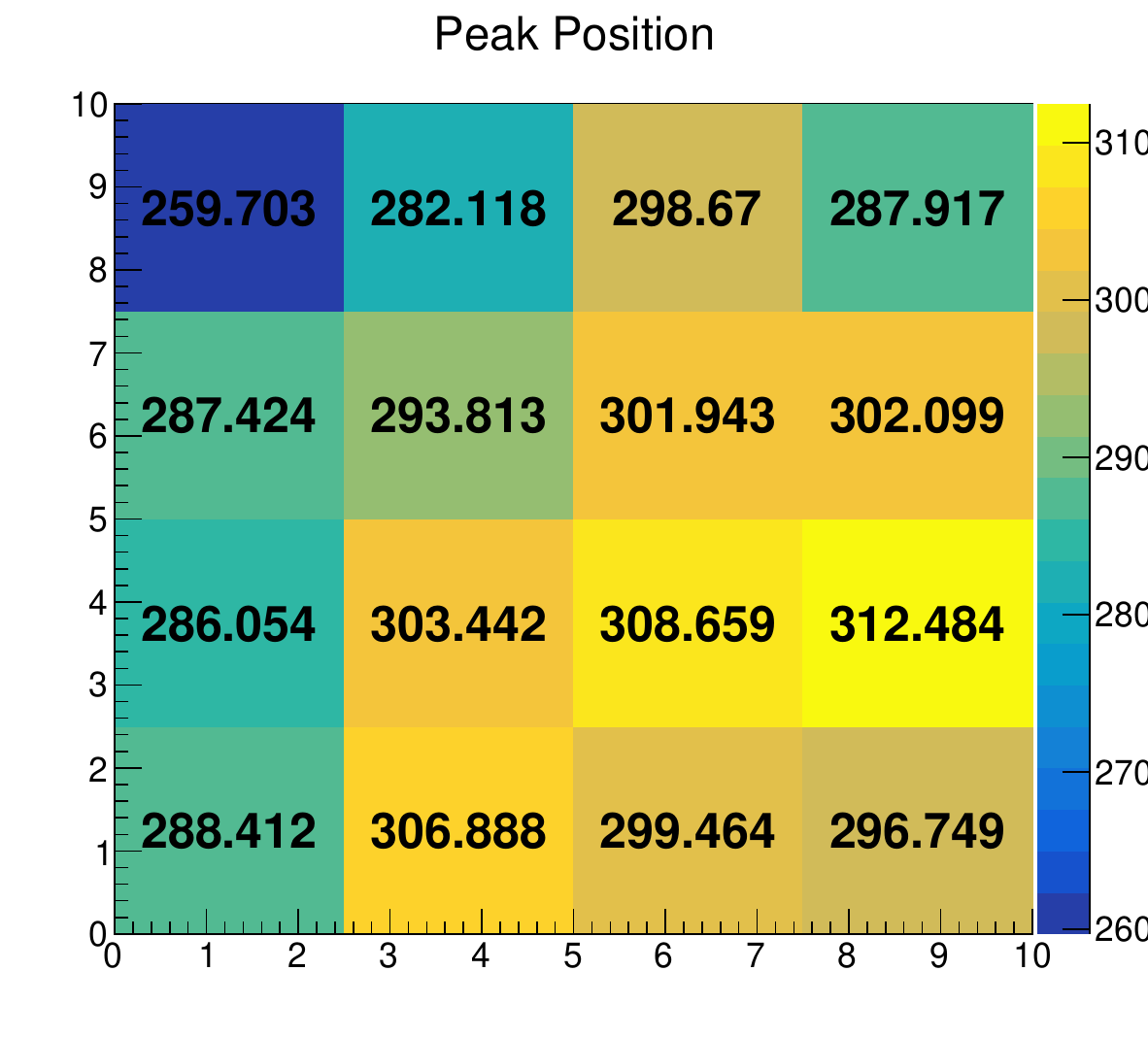}
         \caption{2D Histogram of the position of the peak formed in the spectrum for the whole active area representing the colour map regarding the uniformity of the detector, with the X-ray tube. The values displayed are the position of the peak obtained.}
         \label{fig:pp}
     \label{fig:peakpos}
\end{figure}

Using the setup mentioned previously, one minute of data was acquired for each pixel area and a simple uniformity response map was obtained. The uniformity response of the detector using as reference the peak position as a function of the pixel position is presented in Figure \ref{fig:peakpos}. 
% and \ref{fig:eresuni}.

% For the evaluation of the gain, it was opted not to refer to absolute gain, we only show a relative variation of the peak position in the spectrum, which already gives us an idea on what the gain is (see Equation \ref{eq:gain}), since it makes the calculations simpler and in the final imaging measurements the absolute gain is not obtained. 
To calculate the peak position, a Gaussian fit is applied to each spectrum, from each the amplitude, peak centroid and sigma are retrieved. Figure \ref{fig:peakpos} shows the values of the peak centroid for each pixel. It is possible to notice a strange effect arising from the top left pixel. This smaller peak value can attributed to issues resulting from the assembly of the detector, which resulted in a slight tilt  of the the cathode or \acp{gem} that this way are not full aligned or parallel. Using this values the standard deviation was calculated and found to be 13.2~$\%$, a value inside the standard for this type of detectors \cite{PATRA201725}.

\section{Results and Discussion}
\label{chapt5}

\subsection{Simulation Results}

\begin{figure}[ht]
    \centering
    \includegraphics[scale=0.25]{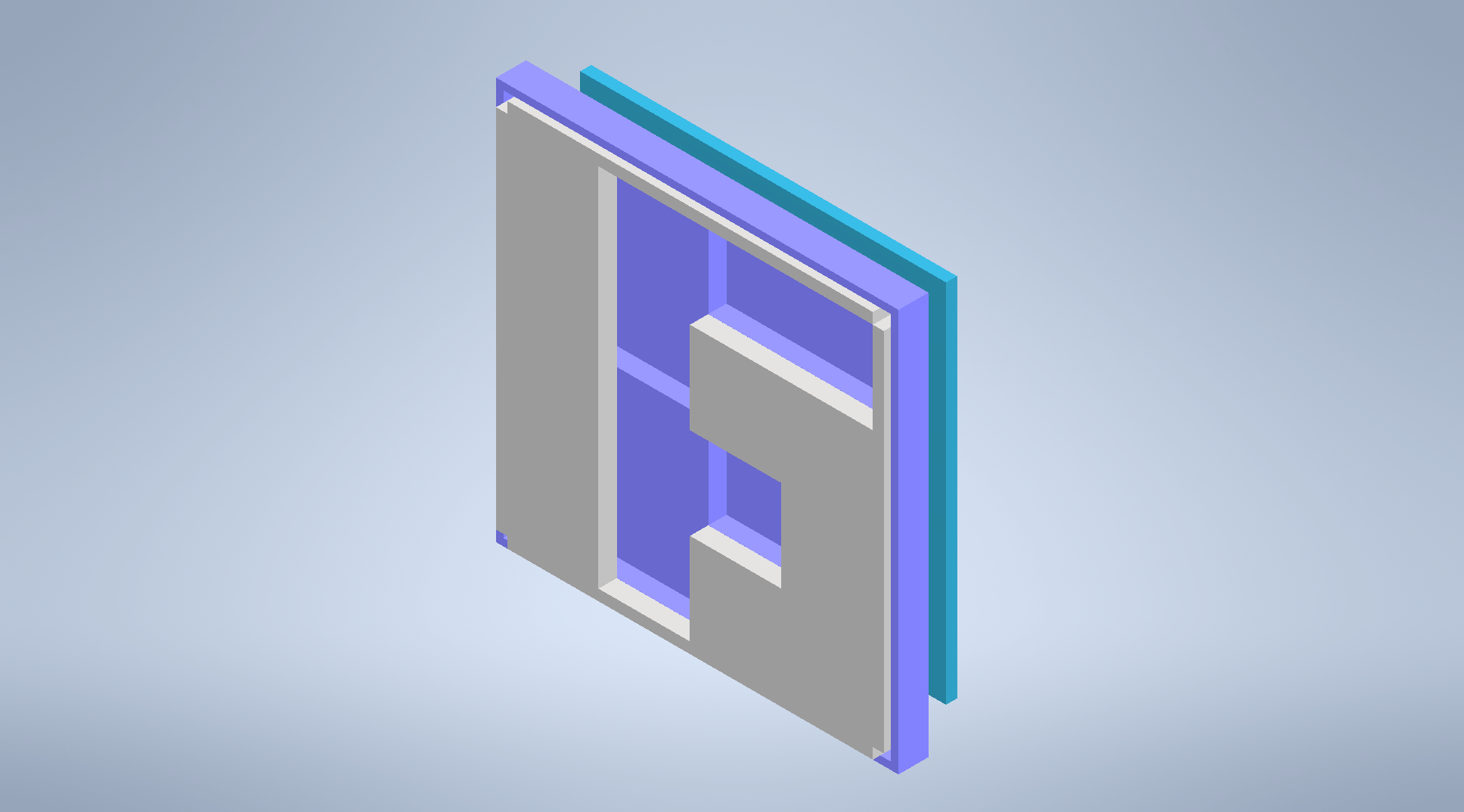}
    \caption{Example of assembly of the setup for the simulations.}
    \label{fig:assembly}
\end{figure}

\begin{figure}[ht]
    \centering
    \includegraphics[width=0.4\textwidth]{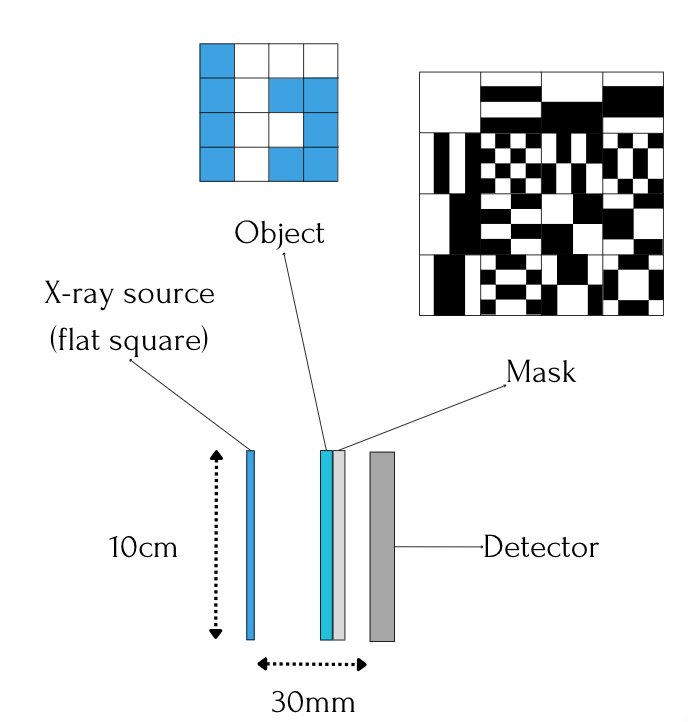}
    \caption{Setup used for the simulations.}
    \label{fig:setupsim2}
\end{figure}

As part of proof of concept, a simulation of the experimental setup was performed using GEANT4. The different components of the setup (sample, masks and detector) were designed using \textit{Autodesk Inventor} and imported to GEANT4, using Ar-CO$_2$ (70-30) at atmospheric pressure for the active volume of detection. Figure \ref{fig:assembly} shows the setup as seen in \textit{Autodesk Inventor}. The setup presented in Figure \ref{fig:setupsim2}, was imported to \textit{GEANT4} using a framework developed at IEAP, that imports the GDML (Geometry Description Markup Language) files converted from the CAD design into \textit{GEANT4} to perform the simulations.

A 10 $\times$ 10 cm$^2$ square uniformly emitting parallel X-rays (flat source) was defined as the particle source, centered accordingly with the rest of the setup and 30 mm away from the detector.  

In the simulation two different types of sources were used according to their energy distributions: monoenergetic and rectangular. For the monoenergetic sources three different energies were simulated: 6, 10, and 15 keV, while the rectangular ones were simulated with a width of 2 keV ranging from 2 to 14 keV to better understand the influence of the different energy ranges in the performance of SPI. The spectrum of the X-ray tube was also used. In addition, a separate set of simulations was performed changing the thickness of the masks for 1, 2, 3 and 4 mm and using the spectrum of the X-ray tube. 

The effect of the different configurations of object plus mask and detector (16 Hadamard masks, plus their inverse) was simulated using 1M events/configuration. The resulting number of counts was then used to reconstruct the image of the object.

\begin{figure}[h!]
    \centering
    \includegraphics[scale=0.7]{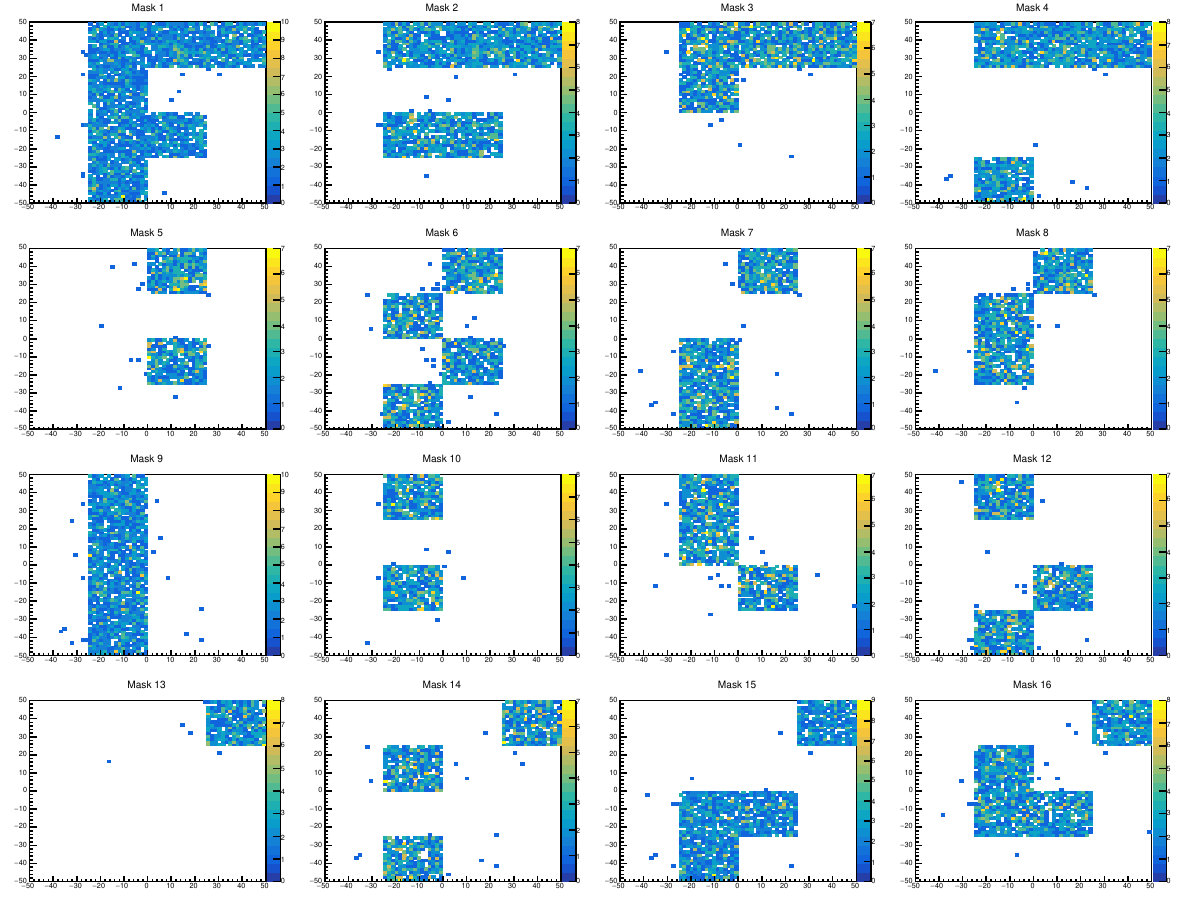}
    \caption{2D histograms obtained for each mask. Example for a monoenergetic source of 6 keV and 5 mm \ac{pla} masks.}
    \label{fig:2dhistograms}
\end{figure}

After obtaining the data for all the patterns, the number of events for each mask was determined by calculating the integral of the energy spectrum, using \textit{ROOT}. A 2D histogram for each mask is shown in Figure \ref{fig:2dhistograms}.

Applying the formula discussed in Section \ref{chap2compal}, it is possible to reconstruct the final image. To evaluate the image quality the \ac{cnr} was calculated for each image using the following equation:

\begin{equation}
    CNR = \frac{signal}{\sqrt{\sigma_s^2 + \sigma_b^2}} = \frac{~\overline{w}-~\overline{d}}{\sqrt{\sigma_w^2 + \sigma_d^2}}
\end{equation}

\noindent where $\sigma_s^2$ and $\sigma_b^2$ are the variances of the signal and the background, respectively and $w$ and $b$, stand for white and dark pixel. In this work, the events registered in the covered/blocked pixels were considered as background, and the signal was determined from the events in unblocked/open pixels. All the values, were first normalized.

\subsubsection{Analysis of the full energy spectrum}

\begin{figure}[h!]
    \centering
    \includegraphics[scale=0.15]{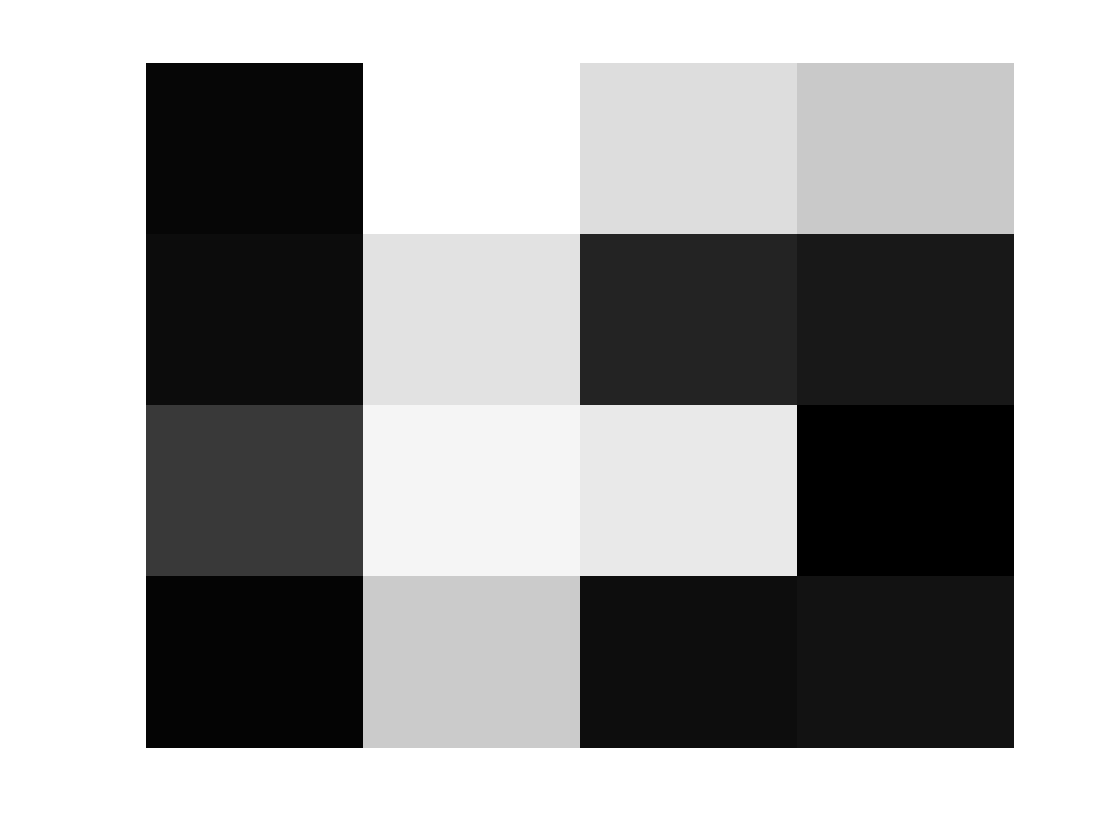}
    \caption{Reconstructed image for the spectrum of the X-ray tube.}
    \label{fig:xray}
\end{figure}

An X-ray source with the same energy distribution of the X-ray used in the experimental component of this work was simulated. Figure \ref{fig:xray} shows the image reconstructed using this approach. 

% \begin{table}[h!]
% \centering
% \caption{\ac{cnr} value obtained for the spectrum of the X-ray tube.}
% \label{tab:contrastxray}
% \begin{tabular}{c|c}
% \cline{2-2}
% \textbf{}           & \textbf{\ac{cnr}} \\ \hline
% X-ray Tube Spectrum & 8.2100              \\ \hline
% \end{tabular}
% \end{table}

Looking at Figure \ref{fig:xray}, it is clearly visible the sample image (letter F). Even though the image is clear, several covered pixels registered events, resulting in a low \ac{cnr} value, about 8.21. This a result of the fact that the source simulated in non-monoenergetic, with the higher energetic X-rays generated being able to reach the active volume of the detector. As it will be discussed next, this can be improved by selecting energy ranges where the X-ray absorption contribution to the fluctuations observed, namely in what concerns the events registered in covered pixels, becomes negligible when compared to other factors.

\subsubsection{Segmented analysis of the spectrum}

\begin{figure}[h!]
  \centering
  \begin{subfigure}[b]{0.32\textwidth}
    \includegraphics[width=\textwidth]{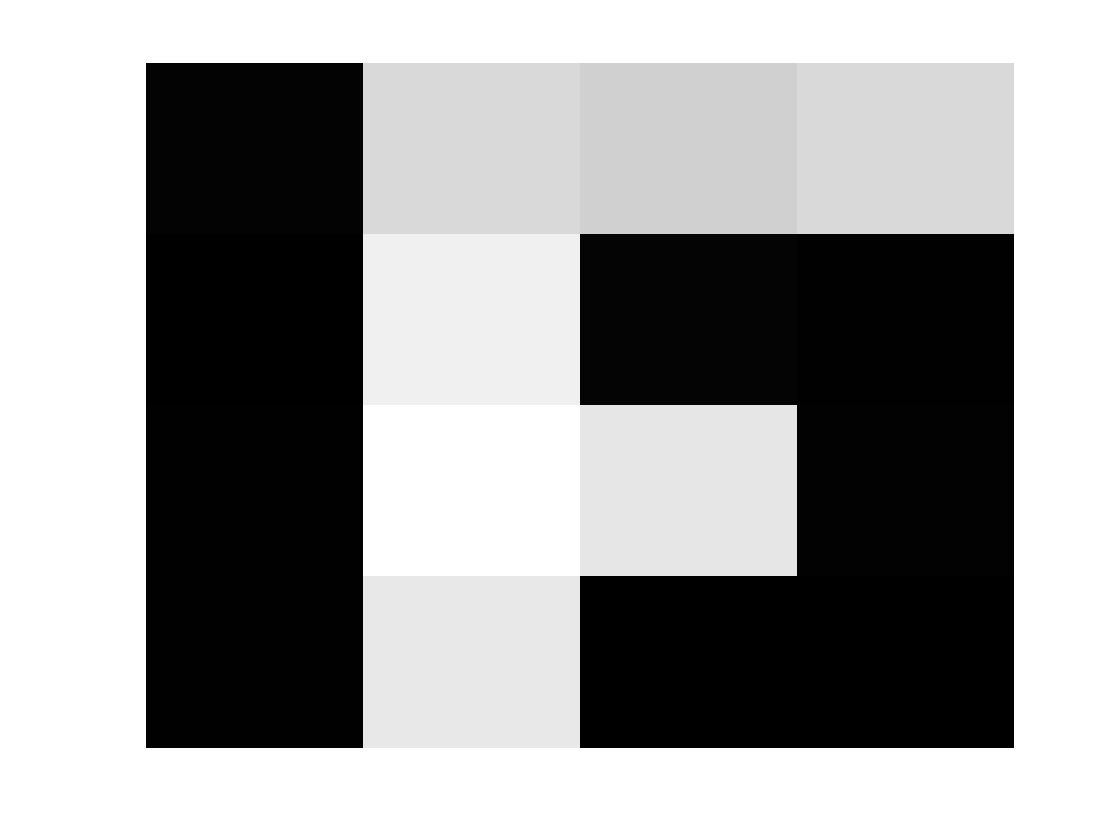}
    % \caption{Reconstructed image for a range from 2 keV to 4 keV.}
    \label{fig:2to4}
  \end{subfigure}
  \hfill
  \begin{subfigure}[b]{0.32\textwidth}
    \includegraphics[width=\textwidth]{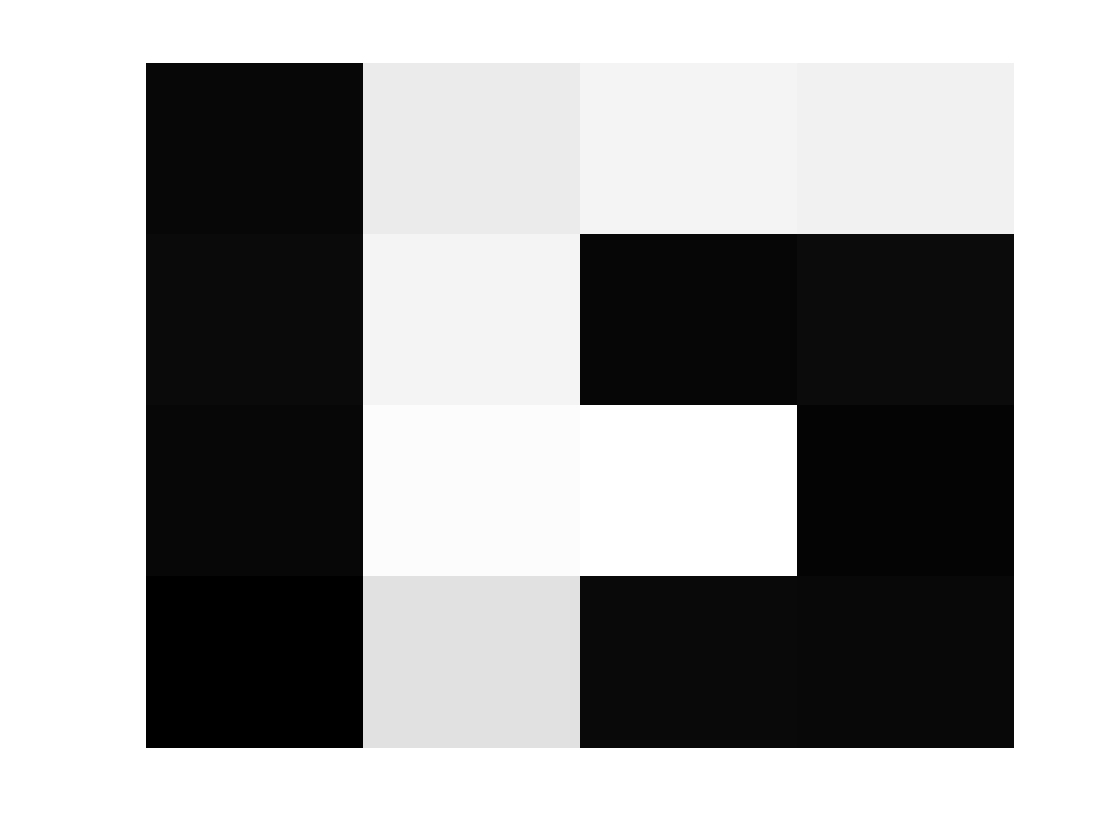}
    % \caption{Reconstructed image for a range from 4 keV to 6 keV.}
    \label{fig:4to6}
  \end{subfigure}
  \hfill
  \begin{subfigure}[b]{0.32\textwidth}
    \includegraphics[width=\textwidth]{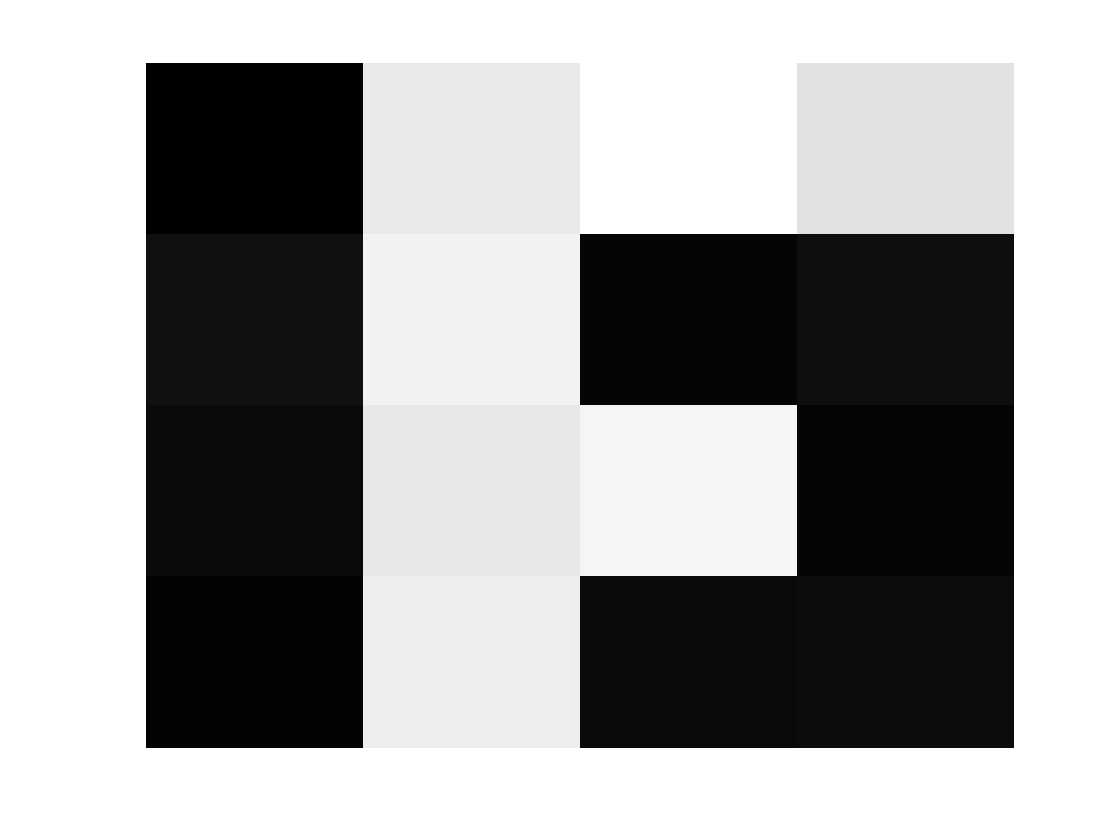}
    % \caption{Reconstructed image for a range from 6 keV to 8 keV.}
    \label{fig:6to8}
  \end{subfigure}
  % \caption{Results for rectangular pulses.}
  \label{fig:pulses1}
  \begin{subfigure}[b]{0.32\textwidth}
    \includegraphics[width=\textwidth]{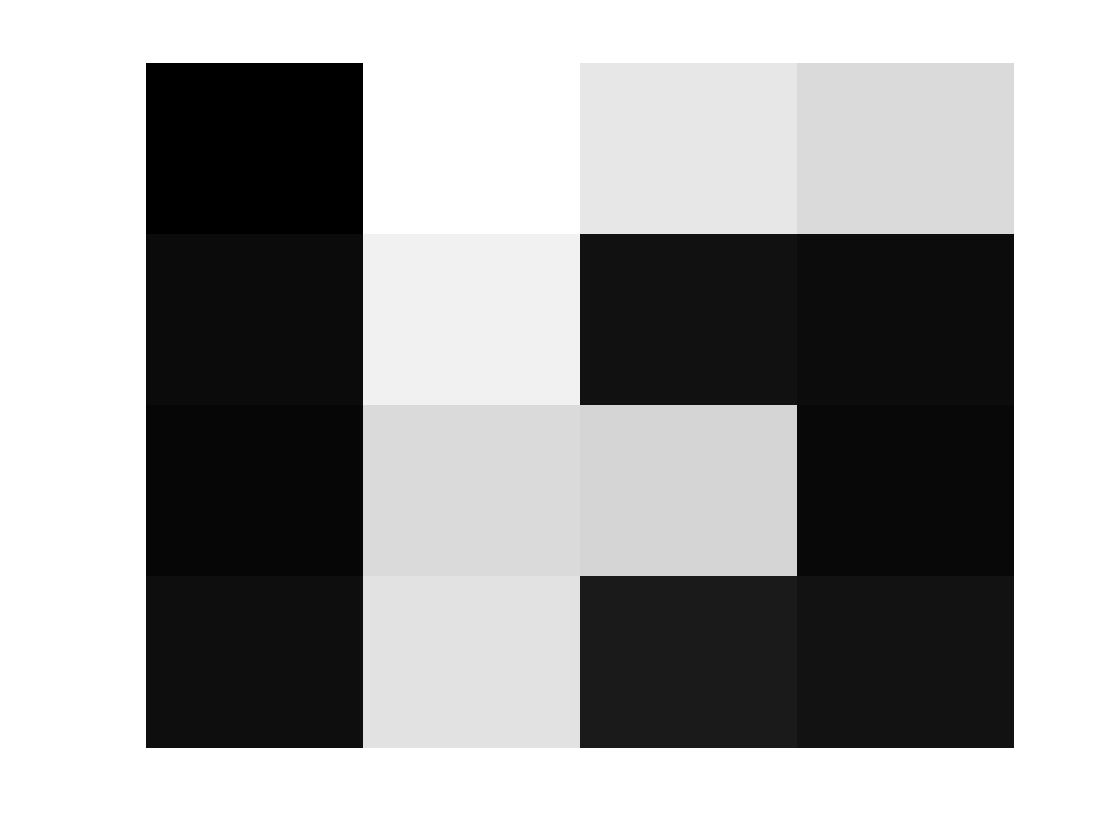}
    % \caption{Reconstructed image for a range from 8 keV to 10 keV.}
    \label{fig:8to10}
  \end{subfigure}
  \hfill
  \begin{subfigure}[b]{0.32\textwidth}
    \includegraphics[width=\textwidth]{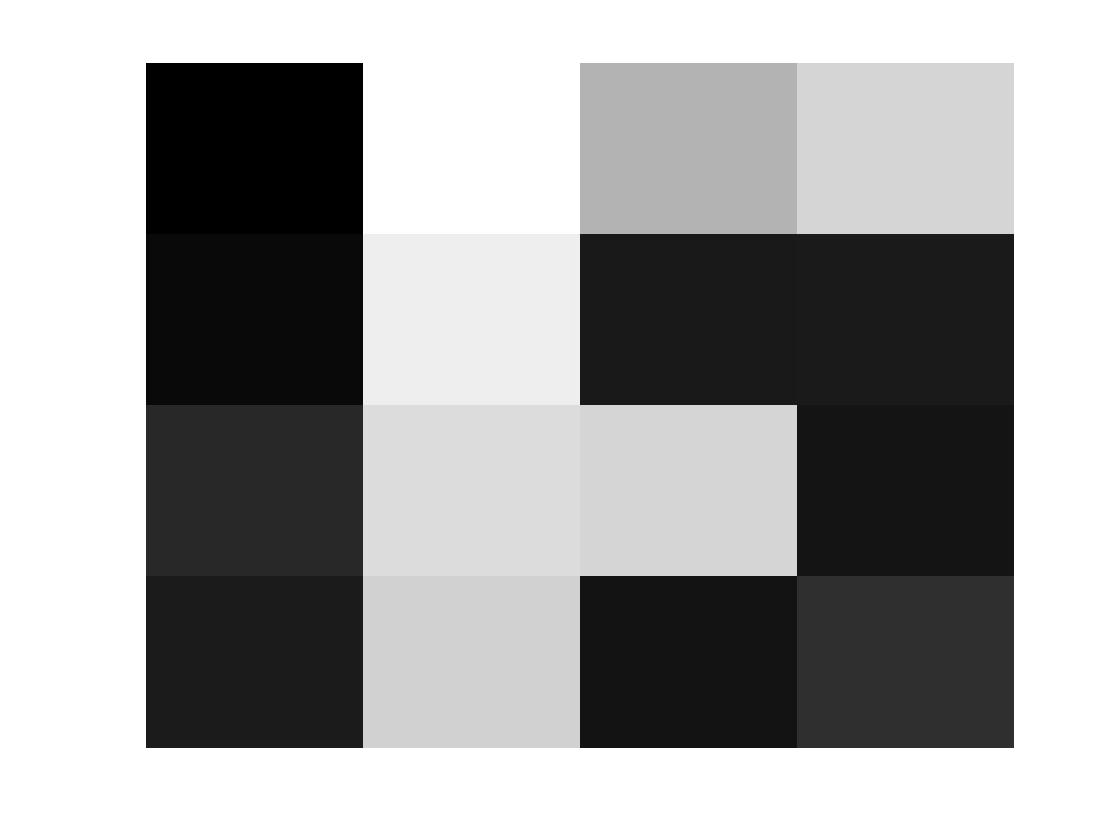}
    % \caption{Reconstructed image for a range from 10 keV to 12 keV.}
    \label{fig:10to12}
  \end{subfigure}
  \hfill
  \begin{subfigure}[b]{0.32\textwidth}
    \includegraphics[width=\textwidth]{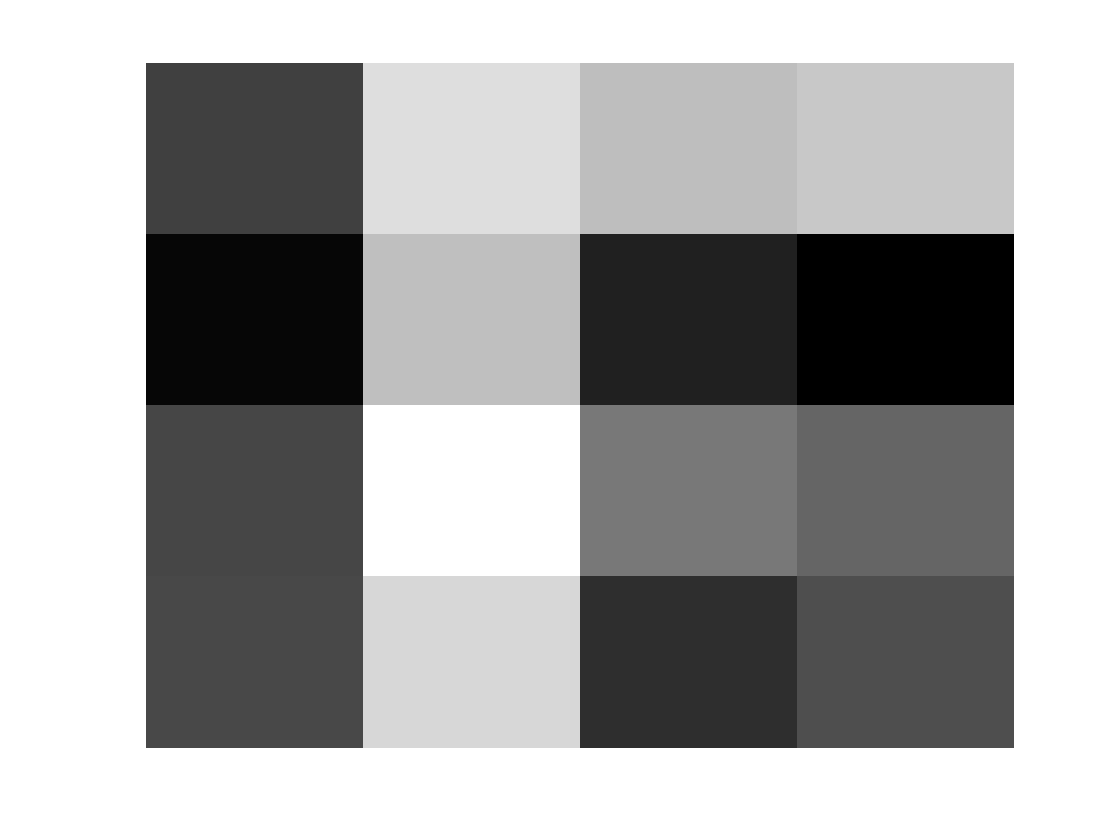}
    % \caption{Reconstructed image for a range from 12 keV to 14 keV.}
    \label{fig:12to14}
  \end{subfigure}
  \caption{Reconstructed image for different energy ranges including a) 2 keV to 4 keV, b) 4 keV to 6 keV, c) 6 keV to 8 keV, d) 8 keV to 10 keV, e) 10 keV to 12 keV and f) 12 to 14 keV.}
  \label{fig:pulses2}
\end{figure}

In order to improve the \ac{cnr} one possibility is to evaluate more convenient energy ranges, which will reduce the number of events in the blocked region that contribute to the background. To do so, different energy ranges were simulated using windows of 2 keV from 2 to 14 keV. In addition to allow a better understanding of the impact of the energy of the X-rays in the image reconstruction, these results will enable a more direct comparison with the experimental results presented later. Figure~\ref{fig:pulses2} shows the reconstructed images for the different energy ranges studied. In table~ \ref{tab:contrastranges} are summarized the respective \ac{cnr} values determined for each image.

\begin{table}[h!]
\centering
\caption{\ac{cnr} values obtained for 2 keV~-wide energy ranges.}
\label{tab:contrastranges}
\begin{tabular}{c|c}
\cline{2-2}
\textbf{X-ray energy range} & \textbf{\ac{cnr}} \\ \hline
2-4 keV   & 15.09           \\ \hline
4-6 keV   & 23.46           \\ \hline
6-8 keV   & 22.20           \\ \hline
8-10 keV  & 13.98           \\ \hline
10-12 keV & 7.47            \\ \hline
12-14 keV & 2.94            \\ \hline
\end{tabular}
\end{table}

Analysing the \ac{cnr} of the images reconstructed, shown in Figure \ref{fig:pulses2}, it is possible to verify that by selecting energy ranges it is possible to improve the image quality (i.e. \ac{cnr}), in particular in the 2-10 keV range. For low energy ranges (2-4 keV), a slightly lower \ac{cnr} was obtained when compared to the performance observed at higher energy ranges (between 4-8 keV). This is something expected as a significant part of these low energy X-rays will be absorbed in the different materials along their path until they reach the active volume of the detector (unblocked), with the detection efficiency increasing above 3 keV. At higher energies, the number of X-ray photons reaching the active volume increases until around 6 keV after which the detection efficiency starts to decrease, as the mask material starts to be transparent to these higher energy X-ray photons. This is a clear effect that results from the intrinsic characteristic of the detector (gas used, density and pressure) as well as the internal configuration (absorption region dimensions), which can be adjusted/changed in order to improve the contrast.
\newpage
% \begin{figure}[h!]
%      \centering
%          \includegraphics[width=0.5\textwidth]{figures/results.png}
%      \caption{Plot of the \ac{cnr} for the different energy sources using the \SI{5}{\milli\meter} \ac{pla} masks. }
%      \label{fig:resultscombinedsources}
% \end{figure}
When simulating mono-energetic sources and comparing with the 2 keV-wide energy ranges a similar trend was observed for X-ray photons with 6 keV. For lower values, in the case of the energy ranges, the \ac{cnr} also diminishes. The results of the simulated full spectrum of the X-ray tube showed a \ac{cnr} value slightly lower than the average of the different 2 keV-wide energy ranges, as its performance is highly affected by the higher energy events generated.

In addition, changing the thickness of the masks resulted in a significant decrease in \ac{cnr} as expected. The \ac{cnr} values obtained are summarized for the different mask thickness in table \ref{tab:contrastthick}. 

Regarding the thickness evaluation, the spectrum of the X-ray tube was used in the simulation. In this case it was possible to conclude that the \ac{cnr} is proportional to the thickness used in the design of the masks, which is consistent with the fact that the X-ray transmission decreases as the thickness increases, eventually saturating for a thickness large enough. The outlier at the thickness of 3 mm,  possibly related with the photon absorption cross section for X-ray photon energies near Argon K-edge (approximately 3.2 keV) that improves the detection efficiency at these energies which are not blocked by the mask for lower thicknesses (below 3 mm) and that around 3 mm are absorbed, eventually with some additional contribution possibly arising from the Cu X-ray fluorescence (7 - 8.5 keV range) (from the GEM electrodes). This will be studied in more detail in the future.

\begin{table}[h!]
\centering
\caption{\ac{cnr} values obtained for different mask thicknesses.}
\label{tab:contrastthick}
\begin{tabular}{c|c}
\cline{2-2}
\textbf{Mask thickness} & \textbf{\ac{cnr}} \\ \hline
5 mm      & 10.66           \\ \hline
4 mm      & 8.96            \\ \hline
3 mm       & 15.96           \\ \hline
2 mm      & 6.32            \\ \hline
1 mm      & 5.45            \\ \hline
0.2 mm     & 3.33            \\ \hline
\end{tabular}
\end{table}

\subsection{Experimental Results}

For the data acquisition, the X-Ray Tube was operated at 10 kV and 5 $\mu$A, and acquisitions were made with different exposure times per mask: 30 seconds, 1 minute, 2 minutes and 5 minutes. In all cases, a .dat file from the \ac{adc} was obtained for each mask, containing a spectrum, 32 in total. The PLA masks used in this work were printed using a Prusa i3 MK1. 

The analysis of the data recorded was performed using \textit{ROOT}. The script used starts by reading the .dat files and calculating a Gaussian fit for each spectrum. The number of events are counted by integrating the energy spectrum. For the determination of the number of events at each measurement the detector gain was considered, allowing to minimize the influence of possible charging-up effects.  
% Figure \ref{fig:rootpart} schematically presents the process described.

% \begin{figure}[h!]
%     \centering
%     \includegraphics[scale=0.55]{figures/rootpart.png}
%     \caption{Scheme of the first step of data analysis: processing and determination of number of events.}
%     \label{fig:rootpart}
% \end{figure}

The same script then makes use of the reconstruction formula presented earlier to calculate the final coefficients for the different energy ranges. The \ac{cnr} values are evaluated for all cases and the best one is selected. Finally, the program returns only the image with best \ac{cnr}, the respective contrast noise and range. 
% Figure \ref{fig:matlabpart} represents a scheme of the program pipeline.

% \begin{figure}[h!]
%     \centering
%     \includegraphics[scale=0.55]{figures/matlabpart.png}
%     \caption{Scheme of the second step of data analysis: Reconstruction and calculation of the \ac{cnr}.}
%     \label{fig:matlabpart}
% \end{figure}

% \subsubsection{SPI Image Reconstruction}

\begin{figure}[h!]
  \centering
  \begin{subfigure}[b]{0.32\textwidth}
    \includegraphics[width=\textwidth]{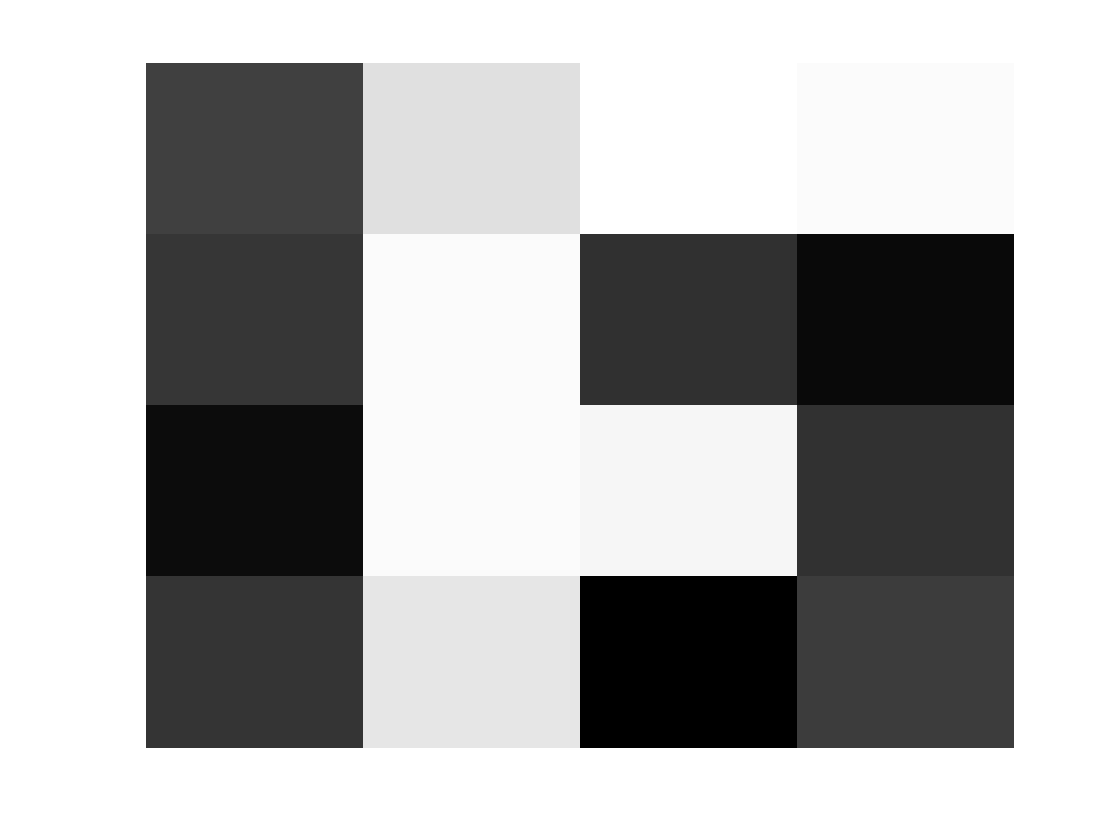}
    % \caption{Reconstructed image for 30 seconds.}
    \label{fig:30sec}
  \end{subfigure}
  \hfill
  \begin{subfigure}[b]{0.32\textwidth}
    \includegraphics[width=\textwidth]{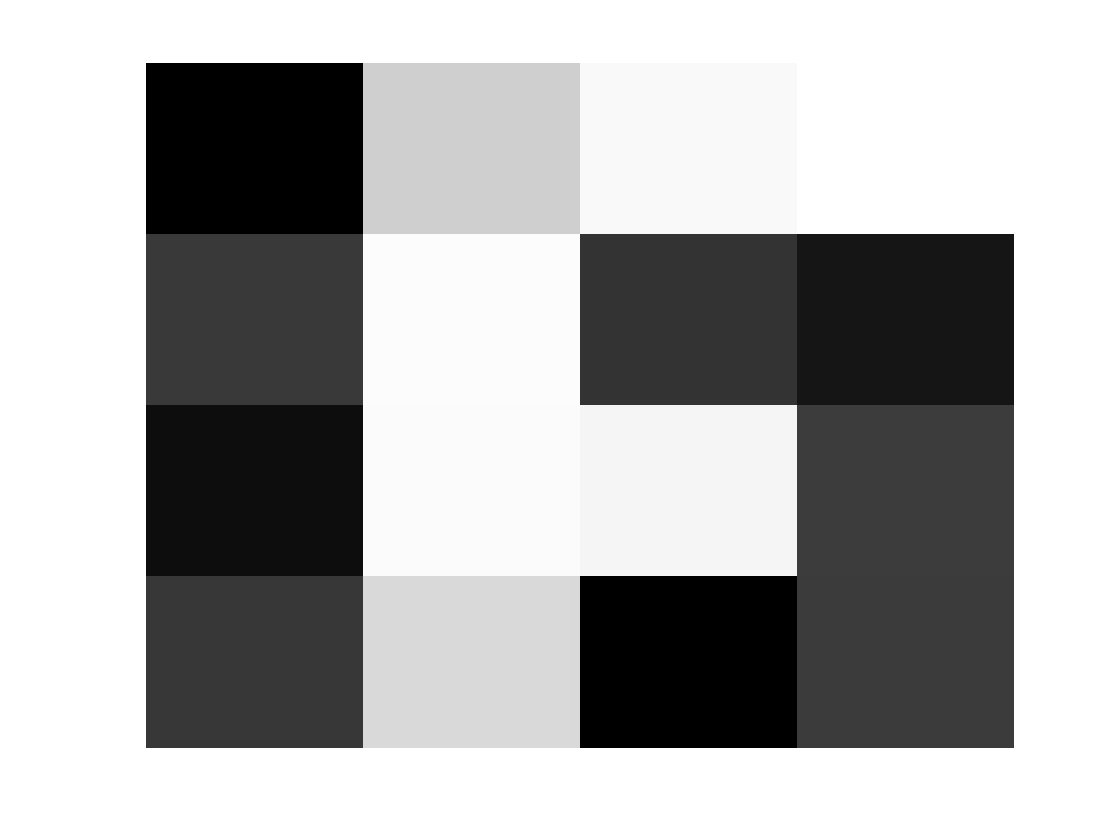}
    % \caption{Reconstructed image for 1 minute.}
    \label{fig:1min1}
  \end{subfigure}
  \hfill
  \begin{subfigure}[b]{0.32\textwidth}
    \includegraphics[width=\textwidth]{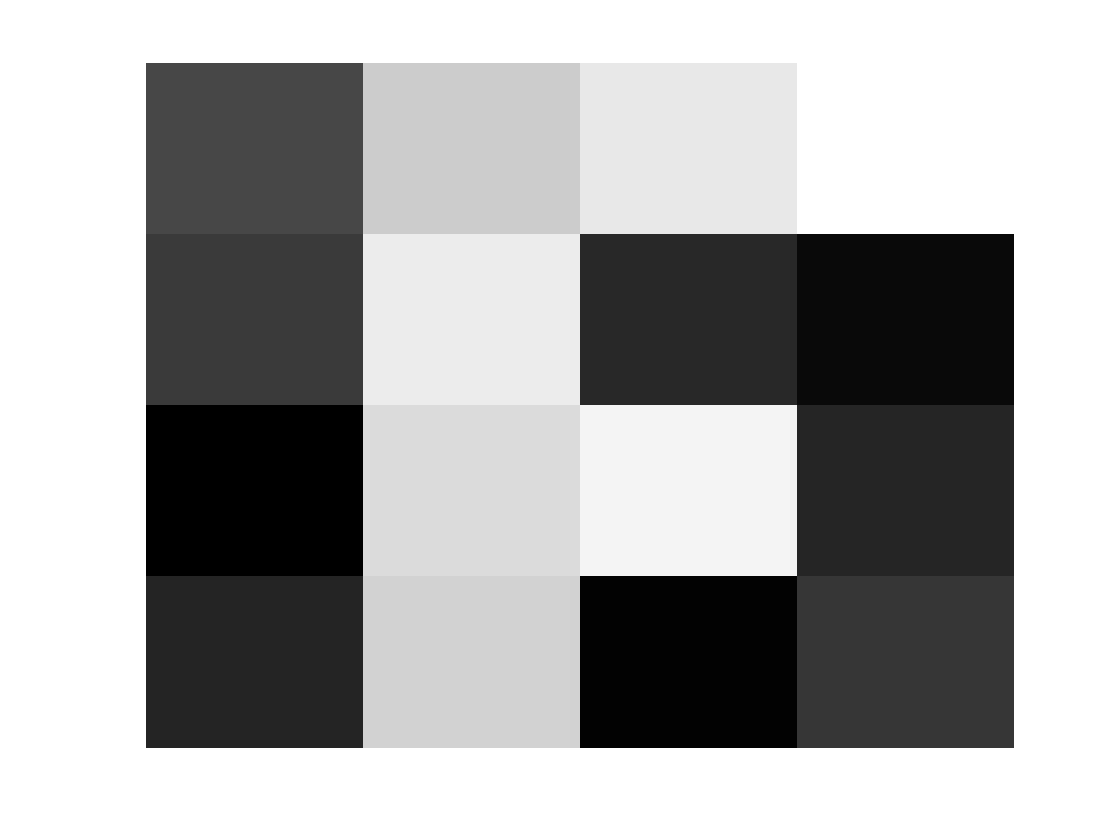}
    % \caption{Reconstructed image for 1 minute.}
    \label{fig:1min2}
  \end{subfigure}
  % \caption{Experimental results for different acquisition times. A) 30 seconds. B) 1 minute. C) 1 minute.}
%   \label{fig:expresults1}
% \end{figure}
% \begin{figure}[h!]
%   \centering
  \begin{subfigure}[b]{0.32\textwidth}
    \includegraphics[width=\textwidth]{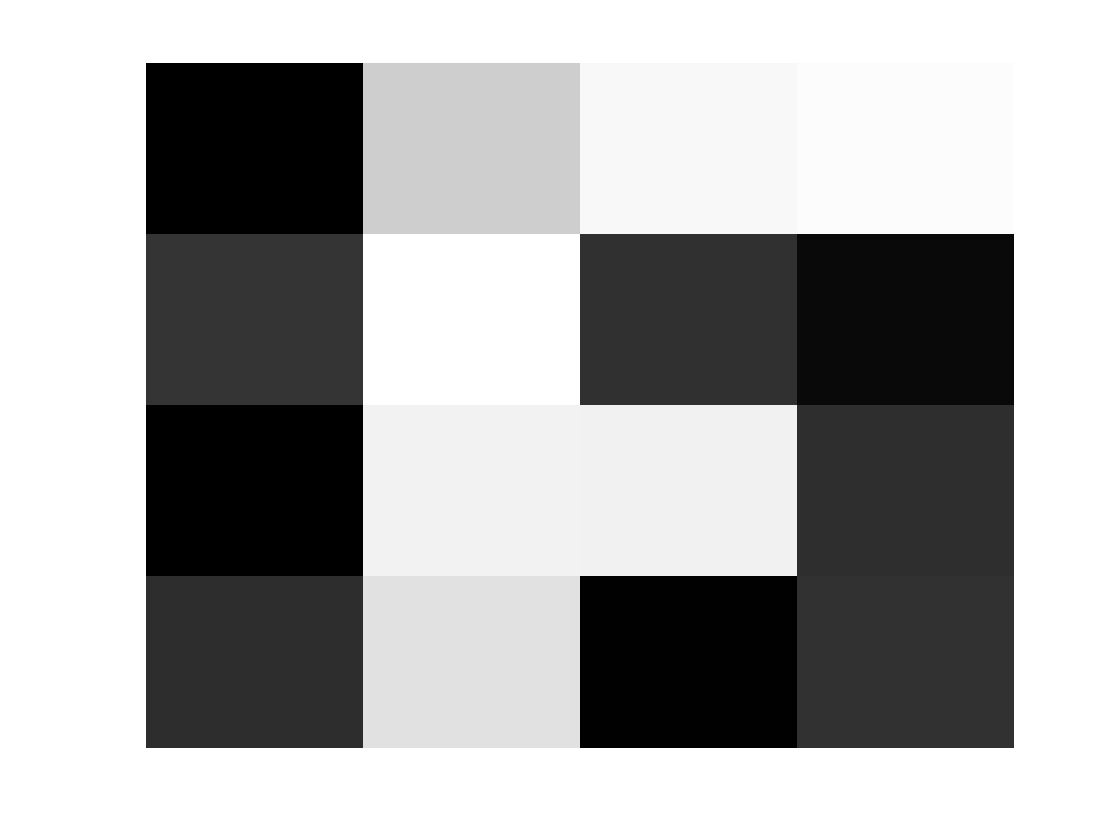}
    % \caption{Reconstructed image for 2 minutes.}
    \label{fig:2min}
  \end{subfigure}
  \hfill
  \begin{subfigure}[b]{0.32\textwidth}
    \includegraphics[width=\textwidth]{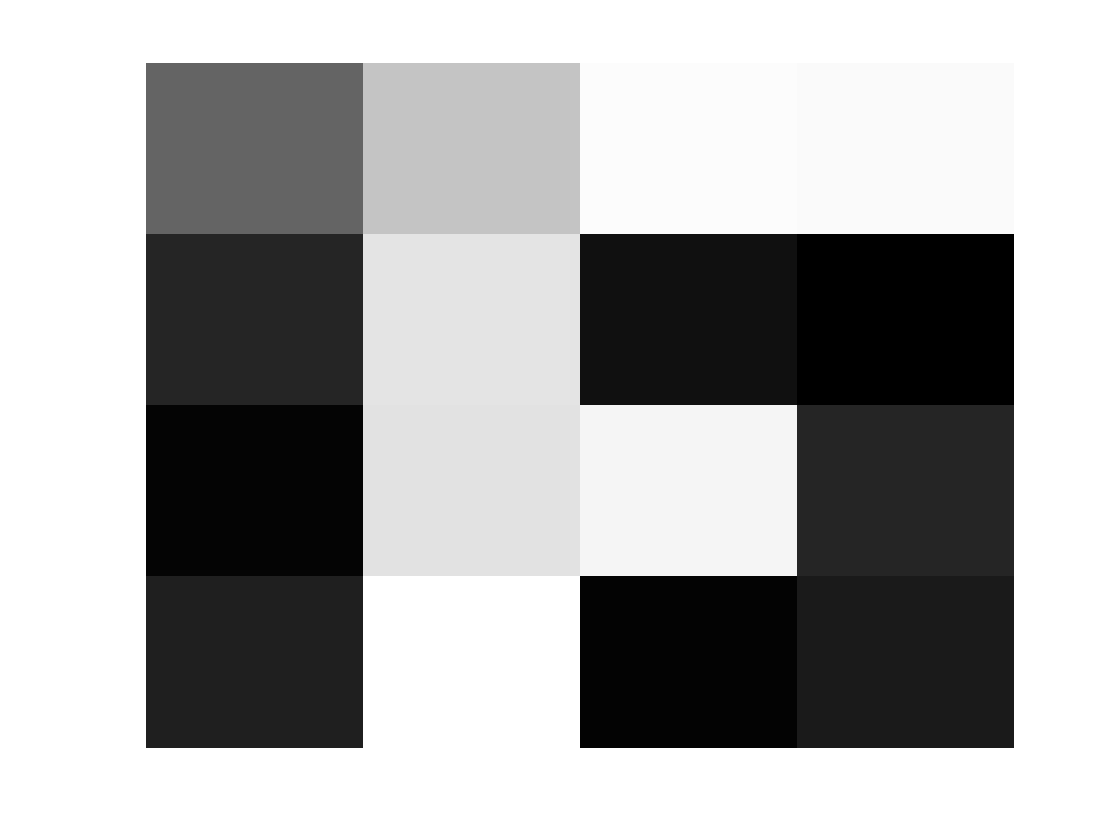}
    % \caption{Reconstructed image for 5 minutes.}
    \label{fig:5min}
  \end{subfigure}
  \hfill
  \begin{subfigure}[b]{0.32\textwidth}
    \includegraphics[width=\textwidth]{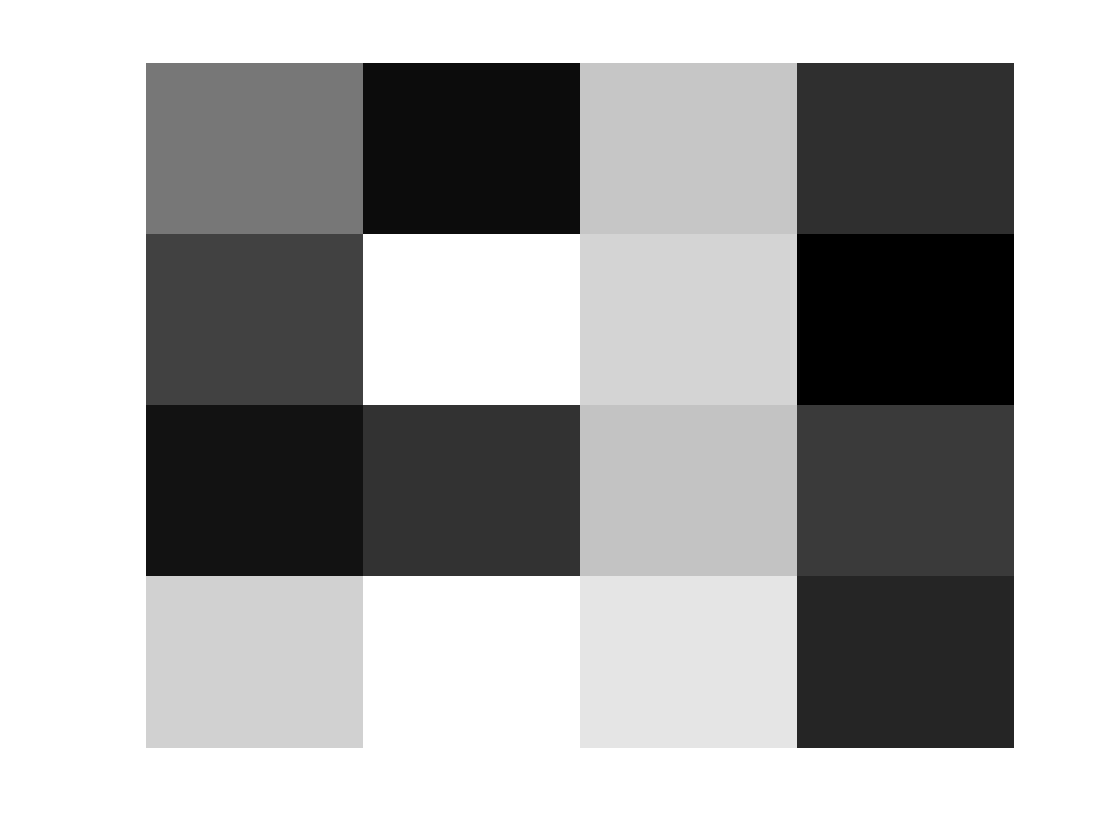}
    % \caption{Reconstructed image for 1 minute with the object inverted.}
    \label{fig:inverted}
  \end{subfigure}
  \caption{Experimental results for different acquisition times and an object inverted. a) 30 s, b) 1 min, c) 1 minutes, d) 2 minutes, e) 5 min and f) 1 minute with the object inverted.}
  \label{fig:expresults2}
\end{figure}

The experimental results obtained are presented in Figure \ref{fig:expresults2} and in table \ref{tab:contrastexp} are summarized the \ac{cnr} values obtained experimentally.
\begin{table}[h!]
\centering
\caption{\ac{cnr} values obtained for the experimental part.}
\label{tab:contrastexp}
\begin{tabular}{c|c}
\cline{2-2}
\textbf{Exposure time} & \textbf{\ac{cnr}} \\ \hline
30 s       & 8.24       \\      \hline
1 min, case 1       & 6.92     \\        \hline
1 min, case 2       & 7.22    \\        \hline
2 min       & 7.60      \\       \hline
5 min       & 8.92       \\      \hline
1 min, object inverted   & 
5.82      \\ \hline
\end{tabular}
\end{table}
%\begin{figure}[h!]
%    \centering
%    \includegraphics[width=0.5\textwidth]{figures/experimentalcnr.png}
%    \caption{\ac{cnr} as a function of the acquisition time per mask of the experimental part.}
%    \label{fig:experimentalresults}
%\end{figure}
The reconstructed images demonstrate the feasibility of the technique, since in all of them it is possible to clearly identify the letter F. These are consistent with the observations reported in Section \ref{secdetuni}. Looking at the \ac{cnr} values obtained experimentally it is possible to see that the variation between the results obtained is small and apparently independent of the exposure time. This may hint that longer exposure times will not necessarily improve the results in terms of contrast, which may be related to possible non-uniformity response of the detector over time. Furthermore, shorter time intervals can still be evaluated in order to establish a minimum limit that still provides enough statistics, since, for 30 seconds, the image has not shown any sign of worsening, having, in fact, one of the best \ac{cnr}, $8.24$. 

\section{Final remarks}

Comparing the simulation and experimental results, it is clear that, even though the experimental results are promising, it was not possible to reach the contrast and image quality obtained in the simulation. Regarding the masks (and object), the 3D printer used is not able to produce a 100$\%$ compact pieces, what would consequently alter the density of the masks, and allow for X-rays to be still transmitted through the material decreasing the calculated contrast. In addition, it is possible that the X-ray photons will not be completely parallel. This was not the case for the simulation, where a flat source (dimensions matching the object and mask size) was used, ensuring that the X-ray photons would impinge the detector perpendicularly to the surface, therefore eliminating this effect. In addition, the features of the detector have to be taken into account, particularly the non-uniformity response, and eventual the rate capability.

\section{Conclusions and future work}\label{sec:xxx5}
We report recent developments in imaging of MPGD-based detectors using SPI techniques, namely the feasibility of X-ray \ac{spi}. The proof of concept was successfully made, both computationally and experimentally.

A functional triple \ac{gem} detector was successfully built and characterized in terms of energy resolution, detector gain and uniformity. The values obtained are in accordance with the literature. A stable gain of around $10^4$ during operation was possible to achieve as planned, corresponding to an energy resolution of around $30 \%$ (5.9 keV FWHM). As mentioned combining MPGD-based detectors with \ac{spi} presents several interesting advantages when compared to competing technologies, as allows to develop versatile solutions for X-ray imaging (flexible design and scalable), cost-effective, these are expected to be ruggedize solutions (radiation hardiness) and capable of withstanding high-rate capability, that with a simplified readout system are able to reconstruction of 2D X-ray  images. In addition to those, it is important to note that combining \ac{gem} with \ac{spi} resulted in a detector immune to the noise caused by the X-ray fluorescence from the copper layers of the \acp{gem}, taking into account the current dimensions. In fact, these photons will artificially increase the efficiency of the detector, by increasing the number of counts in the respective pixel areas.

The results for the simulations of the entire system have shown better results than for the experimental setup, as expected. In fact, to take into account effects related to the detector, such as the non-uniformity resulting from the detector assembly that lead to a position dependent gain, energy resolution; electronic noise, detector charging up effects related to the intensity of the radiation field, among others. A correlation between the \ac{cnr} and the range of energies can be observed in accordance with the expected detector response. The range from 4 to 6 keV demonstrated the best performance, displaying a \ac{cnr} of $23.5$. The acquisition time per mask and the \ac{cnr} did not present a clear proportional relation, due to the statistical fluctuations observed. Although the experimental setup performed well, some aspects can still be improved. Regarding the detector, an optimization of the detector assembly is planned, in order to possibly reduce the observed non-uniformity and minimizing the charging up effect \cite{chargingup}, and improve the performance of the detector setup. 
Potential applications of this technique includes X-ray fluorescence imaging and X-ray imaging through a diffuse media.

\acknowledgments
The authors acknowledge grants MšMT LM2018108 and GAČR GA21-21801S (Czech Republic) and the support from Hugo Natal da Luz and Ali Babar (IEAP/CTU) and for fruitful discussions. A.F.V Cortez has received funding from the European Union’s research and innovation programme: Horizon 2020 under grant agreement No 952480.



%% References


% Bibliography

%% [A] Recommended: using JHEP.bst file
\bibliographystyle{JHEP}
\bibliography{biblio.bib}

%% or
%% [B] Manual formatting (see below)
%% (i) We suggest to always provide author, title and journal data or doi:
%% in short all the informations that clearly identify a document.
%% (ii) please avoid comments such as "For a review'', "For some examples",
%% "and references therein" or move them in the text. In general, please leave only references in the bibliography and move all
%% accessory text in footnotes.
%% (iii) Also, please have only one work for each \bibitem.

\end{document}